\documentclass{aa} 

\usepackage{hyperref} %% to avoid \citeads line fills
\hypersetup{breaklinks=false,
			colorlinks=true,
			citecolor=blue,
			pdfauthor={Luka Poniatowski},
			pdftitle={LTE method},
			pdfsubject={Tables of line-ensemble statistical properties},
			pdfkeywords={Stars: O-stars - Stars: winds, outflows - Stars: mass-loss - Stars: atmospheres},
			pdfproducer={Latex with hyperref},
			pdfcreator={dvips + GPL Ghostscript 9.27}
			}
\renewcommand{\url}[1]{\href{#1}{\texttt{#1}}}% to use the hyphenation in url

\usepackage{txfonts}
\usepackage{multirow}
\usepackage{xcolor}
\usepackage{ulem}

\usepackage{nicefrac}

\usepackage{graphicx}
\usepackage{subcaption}
\graphicspath{{./DMC/}}

\usepackage{natbib}
\bibpunct{(}{)}{;}{a}{}{,} % to follow the A&A style

\usepackage{comment}

\renewcommand{\d}{\mathrm{d}}
\renewcommand{\L}{\mathcal{L}}
\newcommand{\M}{\mathcal{M}}
\newcommand{\R}{\mathcal{R}}
\newcommand{\e}{\mathrm{e}}

\newcommand{\mbf}[1]{\boldsymbol{#1}}

\newcommand{\UnitMd}{\,\mathrm{M_\odot\, yr^{-1}}}
\newcommand{\UnitV}{\,\mathrm{km\,s^{-1}}}
\newcommand{\Unitrho}{\,\mathrm{g\,cm^{-3}}}

\newcommand{\UnitT}{\,\mathrm{kK}}
\newcommand{\Unitkap}{\,\mathrm{cm^2\,g^{-1}}}

\newcommand{\cak}{{\rm CAK}}
\newcommand{\lc}{{\rm LC}}

\newcommand{\eff}{{\rm eff}}

\newcommand{\cl}{{\rm cl}}
\renewcommand{\sb}{{\rm SB}}
\newcommand{\kb}{k_{\rm B}}

\defcitealias{Castor_75}{CAK}
\newcommand{\citecak}{\citetalias{Castor_75}}
\defcitealias{Lattimer_21}{LC}
\newcommand{\citelc}{\citetalias{Lattimer_21}}

\newcommand{\sfrac}[3][1]{    \scalebox{#1}{ $ {}^{#2}{\mskip -5mu/\mskip -3mu}_{#3} $ }   }

\definecolor{coathA_color}{RGB}{22, 82, 145}
\definecolor{coathB_color}{RGB}{16, 110, 23} 
\definecolor{coathC_color}{RGB}{141, 143, 53}
\definecolor{coathD_color}{RGB}{122, 25, 20}
\definecolor{ToDo}{RGB}{205,133,63}
\definecolor{newtxt_color}{RGB}{0, 0, 255}
\definecolor{oldtxt_color}{RGB}{255, 0, 0}

\begin{document}

%\title{New tables of spectral line distribution parameters for opacity computation in acceleration media}
\title{Method and new tabulations for flux-weighted line-opacity and radiation line-force in supersonic media} 
\titlerunning{Tables of line distribution parameters }

\author{L. G. Poniatowski\inst{1} 
            \and 
        N. D. Kee\inst{1,2}
			\and
		J. O. Sundqvist\inst{1}
			\and
		F. A. Driessen\inst{1}
			\and
		N. Moens\inst{1}
			\and
		S. P. Owocki\inst{3}
			\and
		K. G. Gayley\inst{4}
			 \and
		L. Decin\inst{1}
			\and
		A. de Koter\inst{1,5}
			\and 
		H. Sana\inst{1}
	}
\institute{Institute of Astronomy, KU Leuven, 
           Celestijnenlaan 200D, 3001, Leuven, Belgium
		   \and
		   National Solar Observatory, 22 Ohi'a Ku St, Makawao, HI 96768, USA
		   \and
		   Bartol Research Institute, Department of Physics and Astronomy, University of Delaware, Newark, DE19716, USA
		   \and
		   University of Iowa, Iowa City, IA, USA
		   \and
		   Astronomical Institute Anton Pannekoek, Amsterdam University, Science Park 904, 1098 XH Amsterdam, The Netherlands}
              
\date{Received January 12, 2021; accepted April 21, 2022}

 \abstract
{In accelerating and supersonic media, the interaction of photons with spectral lines can be of ultimate importance, especially in an accelerating flow. However, fully accounting for such line forces is computationally expensive and challenging, as it involves complicated solutions of the radiative transfer problem for millions of contributing lines.
This currently can only be done by specialised codes in 1-D steady-state flows. 
More general cases and higher dimensions require alternative approaches.}
{We presented a comprehensive and fast method for computing the radiation line-force using tables of spectral line-strength distribution parameters, which can be applied in arbitrary (multi-D, time-dependent) simulations, including those accounting for the line-deshadowing instability, to compute the appropriate opacities.}
{We assumed local thermodynamic equilibrium (LTE) to compute a flux-weighted line opacity from $\sim4$ million spectral lines. 
We fitted the opacity computed from the line-list with an analytic result derived for an assumed distribution of the spectral line strength and found the corresponding line-distribution parameters, which we here tabulated for a range of assumed input densities $\rho\in[10^{-20},\,10^{-10}]\,{\rm g\,cm^{-3}}$ and temperatures $T\in[10^{4},\,10^{4.7}]\,{\rm K}$.} 
{We found that the variation of the line distribution parameters plays an essential role in setting the wind dynamics in our models. 
In our benchmark study, we also found a good overall agreement between the O-star mass-loss rates of our models and those derived from steady-state studies using more detailed radiative transfer.}
{Our models reinforce that self-consistent variation of the line-distribution parameters is important for the dynamics of line-driven flows. 
Within a well-calibrated O-star regime, our results support the proposed methodology. 
In practice, utilising the provided tables, yielded a factor $>100$ speed-up in computational time compared to specialised 1-D model-atmosphere codes of line-driven winds, which constitutes an important step towards efficient multi-D simulations.
We conclude that our method and tables are ready to be exploited in various radiation-hydrodynamic simulations where the line force is important.}


\maketitle

\section{Introduction}
\label{Se_intro}

Radiation plays a dominant role in the dynamics of various astrophysical environments.
In addition to exchanging energy with the matter via heating and cooling processes, radiation also exchanges momentum with the medium through absorption, emission, and scattering processes. 
Such momentum provided by radiation is dynamically important within, for example, accretion disks around stars, in active galactic nuclei, in cataclysmic variables, and in the atmospheres and winds of stars \citep{Puls_08, Higginbottom_14, Kee_16, Jiang_19}. 

In massive-star atmospheres, the outward push from radiation causes the star to lose mass through powerful radiation-driven stellar winds. 
Prominent spectral features like P-Cygni lines provide observational evidence of radiation-driven wind outflows \citep[for a review see][]{Puls_08, Vink_21}, which already \citet{Lucy_70} theorised are mainly powered by absorption and scattering within spectral lines. 
For a given atom, there can be a vast number of spectral lines for photons to interact with, and in accelerating and transonic media, these effects are further enhanced due to Doppler shifts leading to strong radiation line-forces \citep[see, e.g.,][]{Puls_08, Owocki_15, Sander_20a}.
However, accurately accounting for such `line-driving' in simulations of supersonic flows is a very challenging and computationally expensive problem.

Indeed, state-of-the-art radiation-hydrodynamic (RHD) applications always rely on various approximations when computing the acceleration due to spectral lines. On the one hand, applications where the line force is computed more accurately (i.e. solving the multi-line radiative transfer problem in a reference-frame), have exclusively focused on 1-D steady-state cases \citep[e.g.][]{Sander_17, Sundqvist_19}.
On the other hand, multi-D, time-dependent studies either entirely neglect the dynamic effects of line-driving in supersonic media, that is, they only consider opacities calibrated for static media such as Rosseland mean opacities \citep[e.g.][]{Jiang_18, Jiang_19, Goldberg_21}, or invoke further approximations \citep[e.g.][]{Proga_98, Petrenz_00, Sundqvist_18}. 

In \citet{Poniatowski_21} we proposed a hybrid model wherein Rosseland mean opacities were combined with opacities accounting for line-driving in supersonic regions (as also suggested for the `velocity-stretch' opacity methods discussed by \citet{Castor_04}, chapter 6). For the line-driving, the hybrid opacity model by \citet{Poniatowski_21} relied on a parametrisation first proposed by \citet*{Castor_75} (hereafter \citecak).
They used the so-called Sobolev approximation \citep{Sobolev_60}, assuming that hydrodynamic quantities\footnote{Including atomic level occupation number density and the radiation source function.} are constant over the photon-line-interaction region and approximated the accumulative acceleration from an ensemble of lines as a power-law of the local projected velocity gradient.
\citecak\ computed fitting parameters to their assumed power-law for a set of $C\, {\sc III}$ lines assuming local thermodynamic equilibrium (LTE), and several later studies then updated these fits using more complete lists of spectral lines \citep[e.g.][]{Abbott_82b, Puls_00}.
However, these approaches typically only provide the CAK fit parameters as a function of stellar properties, thus limiting their range of potential applications to (most often one-dimensional and steady-state) simulations of hot-star winds \citep[see also][]{Gormaz_19}.

In time-dependent numerical RHD simulations based a \citecak\ parameterisation, line-force parameters are typically assumed to take some pre-described values; for example, in \citet{Poniatowski_21} an ad-hoc radial stratification for the CAK power-law index parameter $\alpha$ was assumed. 
However, in a more realistic treatment, the parameters describing the line force should rather be self-consistently updated within the simulation itself, according to the local gas and radiation conditions.
A key advantage of our proposed method, e.g. compared to the similar tabulations by \citet*{Lattimer_21} (hereafter \citelc), is that it also offers direct insight into the underlying 
line-distribution function. 
Since most general RHD codes require an opacity as input \citep[see, e.g.][]{Davis_12, Moens_21}, we further choose to work here directly with (flux-weighted) opacities, so that our resulting tables may be used directly in such RHD codes in a way quite analogous to how Rosseland mean opacities (e.g., `OPAL', \citealt{Iglesias_96}) are used in the static limit. 

For example, two applications already underway using our new method regard multi-D extensions of the 1-D Wolf-Rayet (WR) models 
presented in \citet{Moens_21} and 
an extension of the current method to corresponding non-Sobolev, observer's frame line-force calculations (essentially following \citealt{Owocki_96}). This latter will then allow us, for the first time, to investigate effects of self-consistent line-force parameters on the strong `line deshadowing instability' (LDI) \citep{Owocki_84}; thus far, all such LDI simulations have simply used pre-described, prototypical line-force parameters that are assumed to be fixed in time and space \citep[e.g.,][]{Owocki_88, Sundqvist_18, Driessen_21}.    

In this paper, we use the `Munich atomic database' \citep{Pauldrach_98, Pauldrach_01, Puls_05}, which contains over four million spectral lines, to calculate the flux-weighted line opacity (Sect.~\ref{Se_Flux_mean_opacity}). We then fit our results from the complete line-list to an expression derived from the assumed line-strength distribution function, tabulating these by means of our (three) fitting-parameters (Sect.~\ref{Se_Flux_mean_opacity}). To benchmark our method, we apply our tables in a 1-D time-dependent RHD showcase-study of the calibrated grid of line-driven O-star models from \citet{Bjorklund_21}, comparing the resulting wind outflows to predictions for this regime from the literature (Sect.~\ref{Se_first_appl}).
In the final section, we discuss our results and give conclusions (Sect.~\ref{Se_discussion}). 

\section{Flux-weighted mean opacity from spectral lines in an accelerating supersonic medium}
\label{Se_Flux_mean_opacity}
 
As mentioned above, using a statistical approach, one may approximate the flux-weighted mean opacity in both the Sobolev and non-Sobolev limits \citep{Owocki_96}. 
In this paper, we use the Sobolev approximation to evaluate the flux-weighted mean opacity from all contributing lines and then compute the corresponding values of the statistical parameters. Applying this procedure to a range of assumed radiation and gas states, we can then tabulate the resulting parameters as a function of the input state.

\subsection{Flux-weighted mean opacity from a spectral line in the Sobolev approximation}
\label{Se_line_in_Sobolev}

Assuming isotropic emission, the radiation acceleration on a medium illuminated by a frequency-dependent specific intensity $I_\nu$ is:
\begin{equation}\label{Eq_gr_general}
\mathbf{g}_{r} = \frac{1}{\rho c}\int\limits_{0}^{\infty} \oint\d\nu\, \d\Omega \,k_\nu\, \mathbf{\hat{n}}\, I_\nu \, ,
\end{equation}
where $\mathbf{\hat{n}}$ is the unit vector in the direction of radiation propagation, $\rho$ is the mass density in $\Unitrho$, and $c$ is the speed of the light in $\UnitV$. 
The integration is performed for all frequencies $\d\nu$ and over solid angle $\d\Omega = -\d\mu\,\d\phi$, where $\mu = \cos\theta$. 
Furthermore, $k_\nu$ is the extinction coefficient (units of inverse length), related to the mass absorption coefficient $\kappa_\nu$ through $k_\nu = \rho \kappa_\nu$.
For various numerical applications, the radiation acceleration is often formulated using a flux-weighted mean opacity $\bar{\kappa}_{\rm F}$ and radiation flux $\mathbf{F}$ with magnitude $F = |\mathbf{F}|$.
Then $\mathbf{g}_{r}$ simply reads:
\begin{equation}
	\mathbf{g}_{r} = \frac{\bar{\kappa}_{\rm F}\cdot\mathbf{F}}{c}\, ,
\end{equation} 
with:
\begin{equation}
	\label{Eq_flux}
	\mathbf{F} \equiv \int\limits_{0}^{\infty} \oint\d\nu\, \d\Omega \, \mathbf{\hat{n}}\, I_\nu \, .
\end{equation}

A simple comparison of the equations above reveals that the flux-weighted mean opacity is a rank two diagonal tensor:
\begin{equation}
	\label{Eq_kf_general}
	\bar{\kappa}_{\rm F} = 
		\frac{1}{\rho}\frac{\mathbf{F}}{F^2} \int\limits_{0}^{\infty} \oint\d\nu\, \d\Omega \, k_\nu \, \mathbf{\hat{n}}\, I_\nu \, .
\end{equation}

For a single spectral line with line centre frequency $\nu_0$ and $gf$ value corresponding to a specific $l-u$ transition between two atomic levels $l$ (`lower') and  $u$ (`upper'), the extinction coefficient reads: 
\begin{equation}\label{Eq_knu_kl}
k_\nu =  \sigma_{\cl}\;(gf)\left(\frac{n_l}{g_l} - \frac{n_u}{g_u}\right) \;\phi(\nu- \nu_0) = k_L \phi_\nu\, ,
\end{equation}
where $n$ and $g$ are the atomic occupation number density and statistical weight of the respective levels. 
Here, $\sigma_{\cl} = \pi e^2/(m_e c)$ is the classical frequency-integrated line cross-section, with $e$ and $m_e$ being the charge and mass of the electron, and $\phi_\nu = \phi(\nu - \nu_0)$ is the normalised line profile function, with $\nu_0$ the line-centre frequency; $k_L$ then represents the frequency-integrated line-extinction coefficient.  
For such a single line, the components of Eq.~\ref{Eq_kf_general} are: 
\begin{equation}\label{Eq_kf_kL}
	\kappa_{{\rm F\,} (i,\, i)} = \frac{k_L}{\rho}\frac{F_i}{F^2}\int\limits_{0}^{\infty} \oint\d\nu\, \d\Omega \,\phi_\nu \,\hat{n}_i\, I_\nu \,  .
\end{equation}

In general, the specific intensity along some direction $s$ can be computed from a formal solution to the radiative transfer equation 
\begin{equation} 
	\label{Eq_formal_sol}
    I_\nu(\tau_\nu) = I^{\rm core}_\nu\e^{-\tau_\nu} + \int\limits_{0}^{\tau_\nu} S(\tilde{\tau}_\nu)\e^{-|\tau_\nu - \tilde{\tau}_\nu|}\d\tilde{\tau}_\nu= I_\nu^{\rm dir} + I_\nu^{\rm diff} \, ,
\end{equation}
with the line optical depth given by
\begin{equation}
	\label{Eq_tau_line}
	\tau_\nu = \int\limits_{0}^{\infty}\d s \, k_L \phi_\nu\, .
\end{equation}
We have here divided the specific intensity into a direct component $I_\nu^{\rm dir}$, associated with the attenuated specific intensity $I_\nu^{\rm core}$, and a diffuse component $I_\nu^{\rm diff}$, related to the source function $S(\tilde{\tau}_\nu)$ and describing reprocessed radiation. 
Using the Doppler formula, we transform the optical-depth equation above from physical to frequency space. 
Then, under the Sobolev approximation \citep{Sobolev_60}, where the line-of-sight velocity gradient $\d \varv_s/\d s$ and the line extinction $k_L$ are assumed to be constant over the line resonance region, one finds:
\begin{equation}
  \tau_\nu = \frac{k_L c}{\nu_0}\left|\frac{\d \varv_s}{\d s}\right| ^{-1}\int\limits_{0}^{\infty}\d\nu \, \phi_\nu = \tau_S \int\limits_{0}^{\infty}\d\nu \, \phi_\nu \, ,
\end{equation}
where $\tau_S$ is the directional-dependent Sobolev optical depth 
\begin{equation}\label{Eq_tau_sob}
\tau_S(\mathbf{\hat{n}}) =  \frac{k_L c}{\nu_0}\left|\frac{\d \varv_s}{\d s}\right|^{-1}  = \frac{k_L c}{\nu_0\left|\mathbf{\hat{n}}\cdot\nabla(\mathbf{\hat{n}}\cdot\mbf{\varv})\right|}\,.
\end{equation}

For notational simplicity, let us here introduce dimensionless parameters: line strength $q$ \citep{Gayley_95}, an illumination function $w_\nu$, and a characteristic scale $\tau_t$ for the Sobolev optical depth:
\begin{align}%\label{Eq_dimless_params}
&q  = \frac{k_L}{\nu_0 \rho \kappa_0} \label{Eq_dimless_q} \, , \\ 
&w_\nu = \frac{\pi\nu_0\,I^{\rm core}_\nu}{F}\label{Eq_dimless_w} \, , \\ 
&\tau_t(\mathbf{\hat{n}}) = \frac{c\,\rho\kappa_0}{\left|\mathbf{\hat{n}}\cdot\nabla(\mathbf{\hat{n}}\cdot\mbf{\varv})\right|} \label{Eq_dimless_t}\, ,
\end{align}
where $\kappa_0 = 0.34\Unitkap$ is a nominal mass absorption coefficient set to a constant for convenience of the normalisation\footnote{In principle, $\kappa_0$ can be set to any arbitrary value.
However, we choose it to be equal to the mass absorption coefficient of a fully ionised solar plasma, which allows for simple comparisons of our results with earlier studies.}, and using this notation, $\tau_S = q\tau_t$. 
Assuming radially streaming photons from $I_\nu^{\rm core}$ we directly take the integrals in Eq.~\ref{Eq_kf_kL}, where the diffuse component vanishes due to isotropicity.
The radial component of the flux-weighted mean opacity then gives the so-called force multiplier for the single spectral line: 
\begin{equation}\label{Eq_Mt_single}
	M_{\rm line}(\tau_t) = \frac{\kappa_{{\rm F}\, (r,r)}}{\kappa_0} = w_\nu\, q \frac{1 - \e^{-q \tau_t}}{q \tau_t}\,. 
\end{equation}
Note that here the force multiplier is defined for the flux-weighted opacity, rather than the standard definition using the radiation force  (see, e.g. \citecak, \citelc). 

\subsection{Force multiplier for a list of discrete lines}
\label{Se_discrete_sum}

\begin{figure*}%[ht!]
	\centering
		\begin{subfigure}[t]{0.5\linewidth}
			\centering
			\includegraphics[width=1\linewidth]{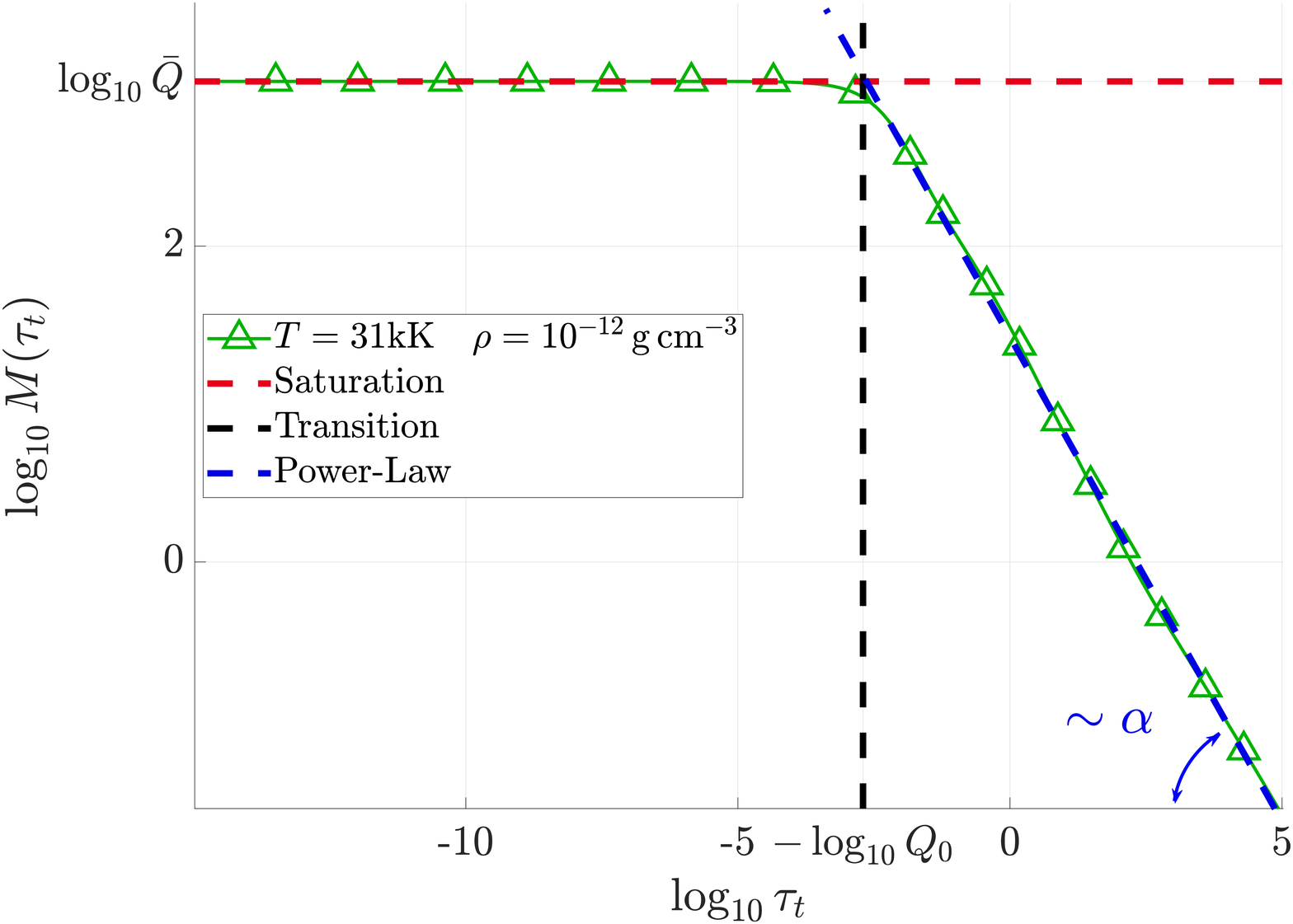}
			\caption{} 
			\label{Fig_Mt_doodle}
		\end{subfigure}%
		\begin{subfigure}[t]{0.5\linewidth}
		   \centering
			\includegraphics[width=1\linewidth]{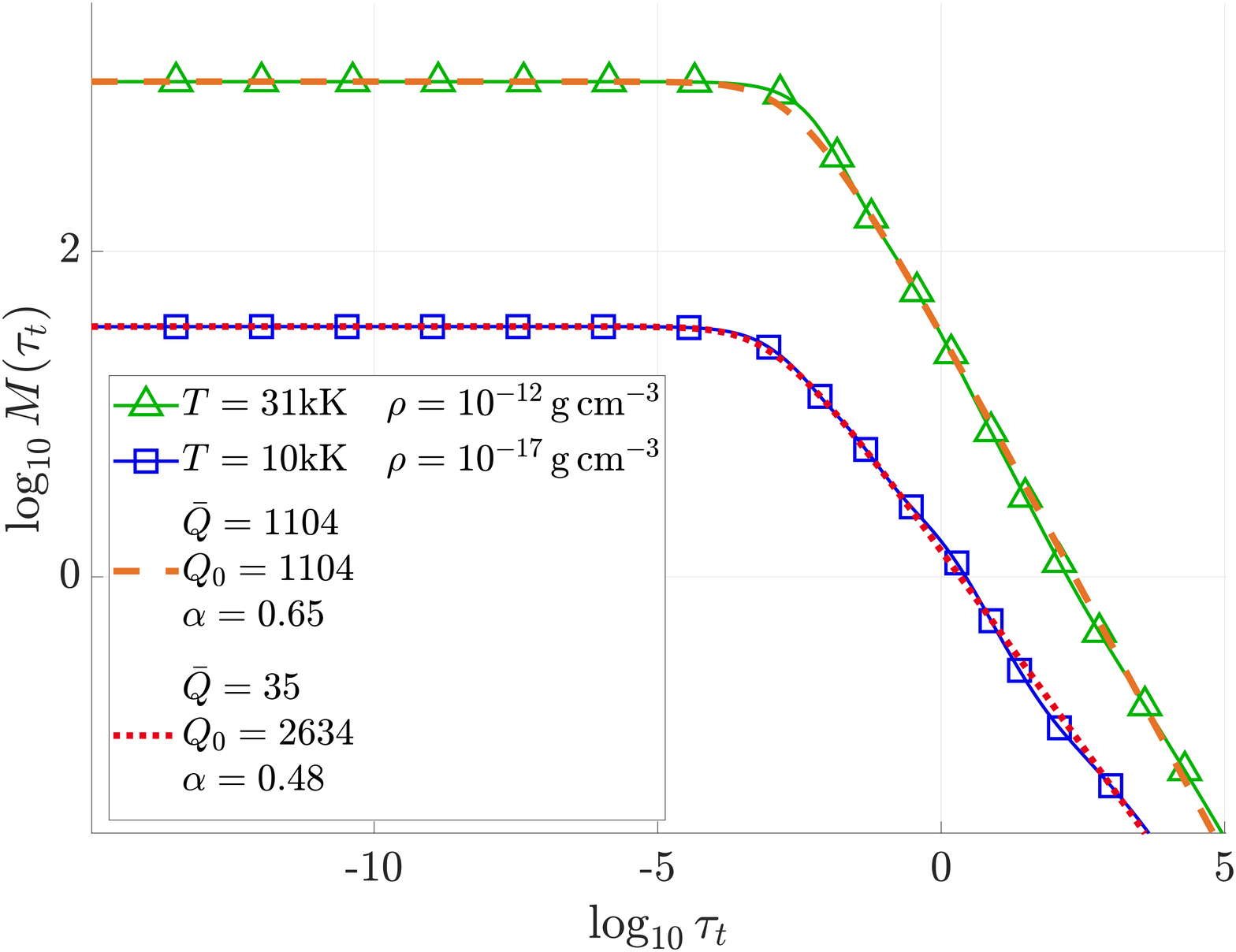}
			\caption{}
			\label{Fig_Mt_fit}
		\end{subfigure}
		\caption{ Force multiplier as computed by evaluating the numerical sum over the discrete set of spectral lines in the triangle-marked green curve.
		Panel (a) shows the saturation level indicated in a red-dashed curve, a power-law region in the blue-dashed line, and the location of a transition point in the black-dashed line. 
		Panel (b) shows the actual fit of two numerical results, where triangle-marked green line has the same temperature and density as in (a) and the square-marked blue line assumes a different set of temperature and density. The derived fit parameters for each case are indicated in the corresponding legend.}
	\end{figure*}

In reality, there can be many thousands of spectral lines contributing towards the full radiation force under various conditions. 
Lists of spectral lines can be found in appropriate databases.
For example, the Munich database we use in this work contains about $4.2\cdot 10^{6}$ lines \citep[for more information see][]{Pauldrach_98, Pauldrach_01}.
For a discrete list of intrinsically non-overlapping lines, we can evaluate the total force multiplier as a sum over all contributing lines \citep[see][\citetalias{Lattimer_21}]{Gayley_95}. 
\begin{equation}\label{Eq_Mt_sum}
M(\tau_t) = \sum\limits_{i = 1}^{\text{all lines}} w_{\nu_i}\, q_i \frac{1 - \e^{-q_i \tau_t}}{q_i \tau_t}\,. 
\end{equation}
Here the subscript $i$ indexes each line in our line list; to evaluate this sum for a given $\tau_t$, we thus need to compute $w_{\nu_i}$, and $q_i$ for each line in our dataset. 

We calculate $w_{\nu_i}$ and $q_i$ using the same approach as \citelc, namely assuming local thermodynamic equilibrium (LTE) to equate the radiation and gas temperatures (i.e, $T_{\rm gas} = T_{\rm rad} = T$).
The incoming intensity and the total flux are further set by the local gas temperature 
via the black-body intensity distribution: 
\begin{align}
	\label{Eq_Icor_plank}
&I^{\rm core}_{\nu_i} = \frac{2h\nu_i^3}{c^2}\cfrac{1}{\e^{\sfrac[1.1]{h\nu_i}{\kb T}}-1}\;,\\
\label{Eq_flux_plank}
&F = \sigma_{\sb} T^4\, ,
\end{align}
where $h$ is the Planck constant, 
$\kb$ the Boltzmann constant, 
$T$ the gas temperature, 
and $\sigma_\sb$ the Stefan-Boltzmann constant.
Combining specific intensity and flux from above in Eq.~\ref{Eq_dimless_w}, we get the illumination function for an unattenuated Planck radiation (i.e., without any spectral features):
\begin{equation}
w_{\nu_i} = \frac{2 h}{\sigma_{\sb}c^2}\frac{\nu_i^4}{T^4}\cfrac{1}{\e^{\sfrac[1.1]{h\nu_i}{\kb T}}-1}\, ,
\end{equation}
where $\nu_i$ is the line-centre frequency of line $i$.

Under this LTE assumption Eq.~\ref{Eq_knu_kl} simplifies to
\begin{equation}
 k_L =  \sigma_{\cl}\;f_{lu}\,n_l\,\left(1 - \e^{{h\nu_i}/{\kb T}}\right)\, ,
\end{equation}
where $f_{lu} = (gf) / g_l$ is the oscillator strength for a given transition. Computing $n_l$ using the Saha and Boltzmann relations (see Appendix~\ref{Se_balance} for details), we also compute $q_i$ using Eq.~\ref{Eq_dimless_q}.

Setting $T$ and $\rho$, we can now evaluate the sum in Eq.~\ref{Eq_Mt_sum} for a given line list and an assumed range of $\tau_t$. 
For $T = 31\,\UnitT$ and $\rho = 10^{-12}\,\Unitrho$, we show the resulting force multiplier for the range $\log_{10} \tau_t = [-15,\,5]$, as a triangle-marked green line on Fig.~\ref{Fig_Mt_doodle}.
Note that $\tau_t$ depends on the ratio of density and velocity gradient rather than on each parameter individually. For a fixed density a range in $\tau_t$ is then equivalent to having a range in velocity gradient. Therefore,  $\tau_t$ can be treated as an independent variable for the purposes of fitting the resulting $M(\tau_t)$ (see also \citelc).
We can directly identify a similar behaviour of the force multiplier as \citelc, characterised by two distinct regimes roughly corresponding to (i) an optically thick ($\tau_t \rightarrow \infty$) region and (ii) an optically thin ($\tau_t \rightarrow 0$) region.
In the optically thick region, $\log_{10} M(\tau_t)$ follows a nearly constant slope as a function of $\log_{10}\tau_t$, which is a well-known result first described by \citecak. 
Following their original notation, the force multiplier there scales with $t$ as
\begin{equation}\label{Eq_Mt_CAK}
M(\tau_t) \propto \tau_t^{-\alpha}\, ,
\end{equation}
where $\alpha$ is a power-law exponent and fit parameter. 
Upon inspection of Fig.~\ref{Fig_Mt_doodle}, it is clear that such a power-law fit, shown as a blue dashed line, describes the optically thick part of the green curve well.
However, as we approach the optically thin part, the force multiplier saturates to a constant value (shown in the Fig.~\ref{Fig_Mt_doodle} with a red-dashed line) instead of continuing to increase following Eq.~\ref{Eq_Mt_CAK}.

We can evaluate the saturation value by taking the optically thin limit $\tau_t \rightarrow 0$ of Eq.~\ref{Eq_Mt_sum}. Then following \citet{Gayley_95}, we define
\begin{equation}\label{Eq_Qb_def}
\bar{Q} = \sum\limits_{i = 1}^{\text{all lines}} w_{\nu_i}\, q_i \, .
\end{equation}
Using $\bar{Q}$, we can now quite accurately describe the force multiplier also in the optically thin parts (neglected by \citecak). 

However, $\bar{Q}$ and $\alpha$ alone are still insufficient to constrain the force multiplier fully. 
For this, we also need to know where the optically thin-to-thick transition occurs, that is, $\tau_t = \tau_{t\,{\rm trans}}$. 
To gain insight into what sets this transition, let us re-examine Eqs.~\ref{Eq_Mt_sum} and \ref{Eq_Qb_def}, noting that saturation can only take place for the majority of lines when 
\begin{equation}
	\frac{1 - \e^{-q_i\,\tau_t}}{q_i\,\tau_t} \la 1\, .
\end{equation}
Upon closer inspection of Fig.~\ref{Fig_Mt_doodle}, we see that the saturation occurs already at some $\tau_{t\,{\rm trans}} > 0$.
In other words, the line strength is limited to some effective maximum, which we define as $Q_0$ \citep[see also,][]{Owocki_88}. 
We can roughly estimate this maximum strength by locating the intersection of the power-law at high optical depth with the constant saturation level at low optical depth, such that $Q_0 \approx 1/\tau_{t\,{\rm trans}}$; in Fig.~\ref{Fig_Mt_doodle} we highlight the intersection with a black-dashed line.

\subsection{Force multiplier fitted to an exponentially truncated power-law line distribution}
\label{Se_Power_law}

We can now calculate the force multiplier for various different combinations of temperature and density.
For each combination of $(T,\,\rho)$ appropriate values of $\alpha$, $\bar{Q}$, and $Q_0$ can be tabulated and later used to reproduce the value of the force multiplier without using the spectral line databases. 
This approach is quite analogous to \citelc, differing mainly by our choices of fitting parameters to the force multiplier. 
However, a key advantage with the parametrisation here is that it allows us to capture (and analyse) the underlying statistical distribution of spectral lines. 
As mentioned in Sect.~\ref{Se_intro}, applications like LDI require this spectral line-distribution function to be specified in order to compute the corresponding non-Sobolev radiation line force \citep[][]{Owocki_96}, which as we discuss in Sect.~\ref{Se_disc_comp_to_lc} will now be possible using our tabulations of line strength parameters tabulated here  

To provide insights into the statistical properties of the spectral line distribution, it is instructive to start by assuming an underlying distribution for the number of lines across the frequency and line-strength domains and then to investigate the resulting force multiplier.
For the functional form of the line distribution, we use a parametrisation suggested by \citet{Gayley_95}, where the number of lines $\d N$ in some interval of strength $(q,\;q+\d q)$ and frequency interval $(\nu,\;\nu+\d\nu)$ is:
\begin{equation}\label{Eq_dN}
\d N_\nu = f(\nu)\frac{\bar{Q}}{\Gamma(\alpha)\,Q_0^2}\left(\frac{q}{Q_0}\right)^{\alpha - 2} \e^{-{q}/{Q_0}}\;\d\nu\;\d q\,.
\end{equation}
Here, $\Gamma(\alpha)$ is the Gamma function, and $f(\nu)$ describes the frequency distribution of line strength that we assume to follow $f(\nu)  = 1/\nu$ \citep[see also][]{Friend_83,Puls_00}.
Combining this with Eq.~\ref{Eq_Mt_single}, we integrate over the total number of lines in frequency range 
%($\nu\in[0,\,\infty)$) 
and line strength: %($q\in[0,\,\infty)$):
\begin{equation}\label{Eq_Mt_Gayley}
	M_{G}(\tau_t) = \iint\limits_{0}^{\infty} \d N_\nu\, w_\nu\, q \frac{1 - \e^{-q \tau_t}}{q \tau_t}  =  \cfrac{\bar{Q}}{1-\alpha}\cfrac{(1+Q_0\tau_t)^{1-\alpha}-1}{Q_0\tau_t}\,, 
\end{equation}
where the subscript $G$ refers to \citeauthor*{Gayley_95}.
This description has several advantages as compared to alternatively proposed parametrisations \citep[see also][]{Gayley_95}.
First, it leads to a simple analytic form of the force multiplier that is sufficient to reproduce the observed asymptotic behaviour found for a discrete set of spectral lines.
We can verify this by simply taking the optically thick and thin limits in Eq.~\ref{Eq_Mt_Gayley}.
In the optically thick limit, when $Q_0 \tau_t \gg 1$, we directly obtain the same power-law dependency as in Eq.~\ref{Eq_Mt_CAK}
\[M_{G}(\tau_t)\approx \frac{\bar{Q}Q_0^{-\alpha}}{1-\alpha} \tau_t^{-\alpha}\, , \]
such that when using the \citet{Gayley_95} Ansatz  $Q_0 = \bar{Q}$, we recover the well-known form of the \cak\ force multiplier:
\[M_{G}(\tau_t) = \frac{\bar{Q}^{1-\alpha}}{1-\alpha} \tau_t^{-\alpha}\, .\]
In the opposite limit $Q_0 \tau_t \ll 1$, a first-order Taylor expansion of Eq.~\ref{Eq_Mt_Gayley} gives:
\[M_{G}(\tau_t)  \approx \bar{Q}\, .\]

This means that, for each combination of $(T,\rho)$ used to compute $w_{\nu_i}$ and $q_i$ for all lines in our database, Eq.~\ref{Eq_Mt_Gayley} can be applied to fit $M(\tau_t)$ resulting from Eq.~\ref{Eq_Mt_sum} across a very wide range of $\tau_t$, covering both the optically thick and thin limits, and using only the three basic fit-parameters $\bar{Q}$, $Q_0$, and $\alpha$.
Second, this parametrisation offers insight into the underlying distribution of spectral lines through intuitive parameters.
As in Sect.~\ref{Se_discrete_sum}, $\bar{Q}$ and $Q_0$ correspond to the maximum force in the optically thin limit and the effective maximum strength of spectral lines, respectively. 
The values of the $\alpha$, $\bar{Q}$, and $Q_0$ parameters can now be found by directly fitting Eq.~\ref{Eq_Mt_Gayley} to the actual force multiplier as computed from the discrete line list. 
Note that here we do not consider the so-called $\delta$ parameter, introduced by \cite{Abbott_82b} to account for the shift of ionisation states over the wind.
In cases when the line distribution parameters are held fixed throughout the full wind, the $\delta$ parameter then accounts for the fact they should vary with position. 
This is typically applied in studies of 1-D line-driven stellar winds, where the line distribution parameters have often been tabulated as a function of stellar parameters \citep[i.e., effective temperature, mass and radius of the star,][]{Abbott_82b}, and each considered wind model thus has been characterised by one single value for the line distribution parameters.
Those models then require the $\delta$ parameter to introduce back the radial variation neglected by the use of spatially fixed line force parameters \citep[see also][]{Gormaz_19}. 
In our new method presented here, these effects are instead directly accounted for by allowing the line-distribution parameters to vary with the local density and temperature (thus, they also vary naturally with both position and ionisation), see a first benchmark application in Sect.~\ref{Se_first_appl}.

Figure~\ref{Fig_Mt_fit} shows the results of such a fit for two different combinations of temperature and density. 
Here, the triangle-marked green curve is the same force multiplier as in Fig.~\ref{Fig_Mt_doodle} and the square-marked blue line is the force multiplier for $T = 10\,\UnitT$ and $\rho = 10^{-17}\,\Unitrho$.
The golden-dashed and red-dotted curves are the corresponding best-fitting $M_{G}(\tau_t)$.
Though we note that this parametrisation might not be sufficient to fully capture some of the fine details of the optically thick region, as well as perfectly reproduce the force multiplier in the transition region \citepalias[see discussion in][]{Lattimer_21}, it is sufficient to represent the overall character of the force multiplier while also providing insight on the spectral line distribution. 
As shown explicitly in Sect.~\ref{Se_Model-0}, the range in $\tau_t$ over which we compute $M(\tau_t)$ is sufficient to cover the full range of values present in our simple 1D RHD benchmark simulations of O-star winds.
Moreover, since for a given $(\rho,T)$ pair $M(\tau_t)$ saturates at $\tau_t\ll\tau_{t\,{\rm trans}}$ and follows a single power-law for $\tau_t\gg\tau_{t\,{\rm trans}}$, fitting over an even larger range of $\tau_t$ would in any case not yield meaningfully different values for the fit parameters.
The specific values for density and temperature showcased in the triangle-marked green curve in Fig.~\ref{Fig_Mt_fit} are rather good representations of the expected values for the already quite extensively analysed regime of early O-supergiants.
Our best fit values $\alpha=0.65$, $\bar{Q}= Q_0 = 1104$ are indeed broadly consistent with previous studies of this region \citep{Gayley_95, Puls_00}. 
In contrast, for the second pair of density and temperature, which is outside the early O-supergiant regime, the best fit parameters $\alpha=0.48$, $\bar{Q}=35$ and $Q_0=2634$ are very much outside these previously estimated ranges. 
This illustrates a general need for not assuming constant line-force parameters independent of conditions, which so far often has been the case in (at least time-dependent) simulations of line-driven flows. 
Let us thus next expand the calculations above towards a much broader range of temperature and density combinations.

\subsection{Grid of line distribution parameters}
\label{Sec_parameter_tables}

\begin{figure*}[ht!]
	\centering
		\begin{subfigure}[t]{0.5\linewidth}
			\centering
			\includegraphics[width = 0.95\linewidth]{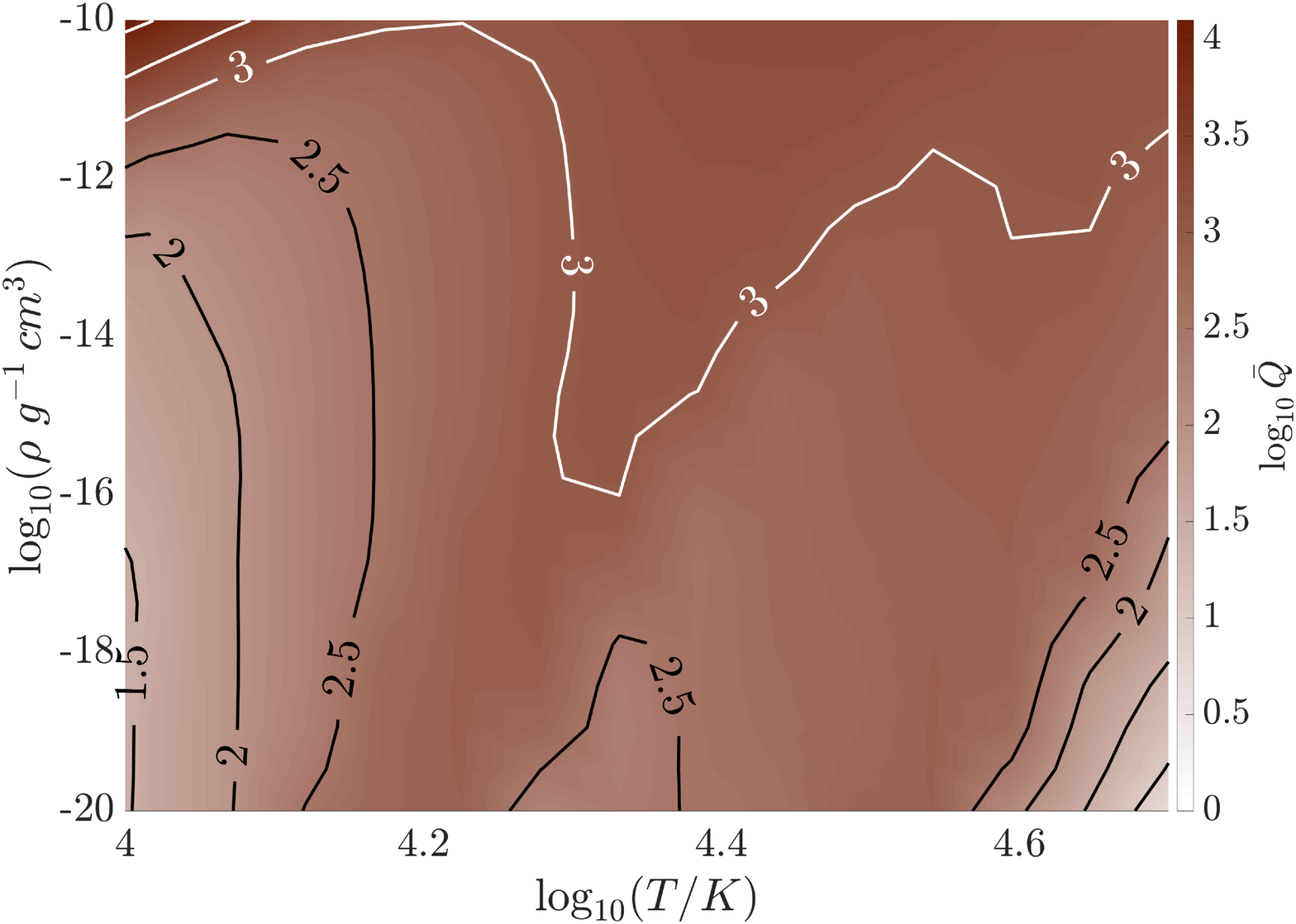}
			\caption{}
			% \label{Fig_map_ke}
			\label{Fig_map_qb}
		\end{subfigure}\hfill
		\begin{subfigure}[t]{0.5\linewidth}
			\centering
			\includegraphics[width = 0.95\linewidth]{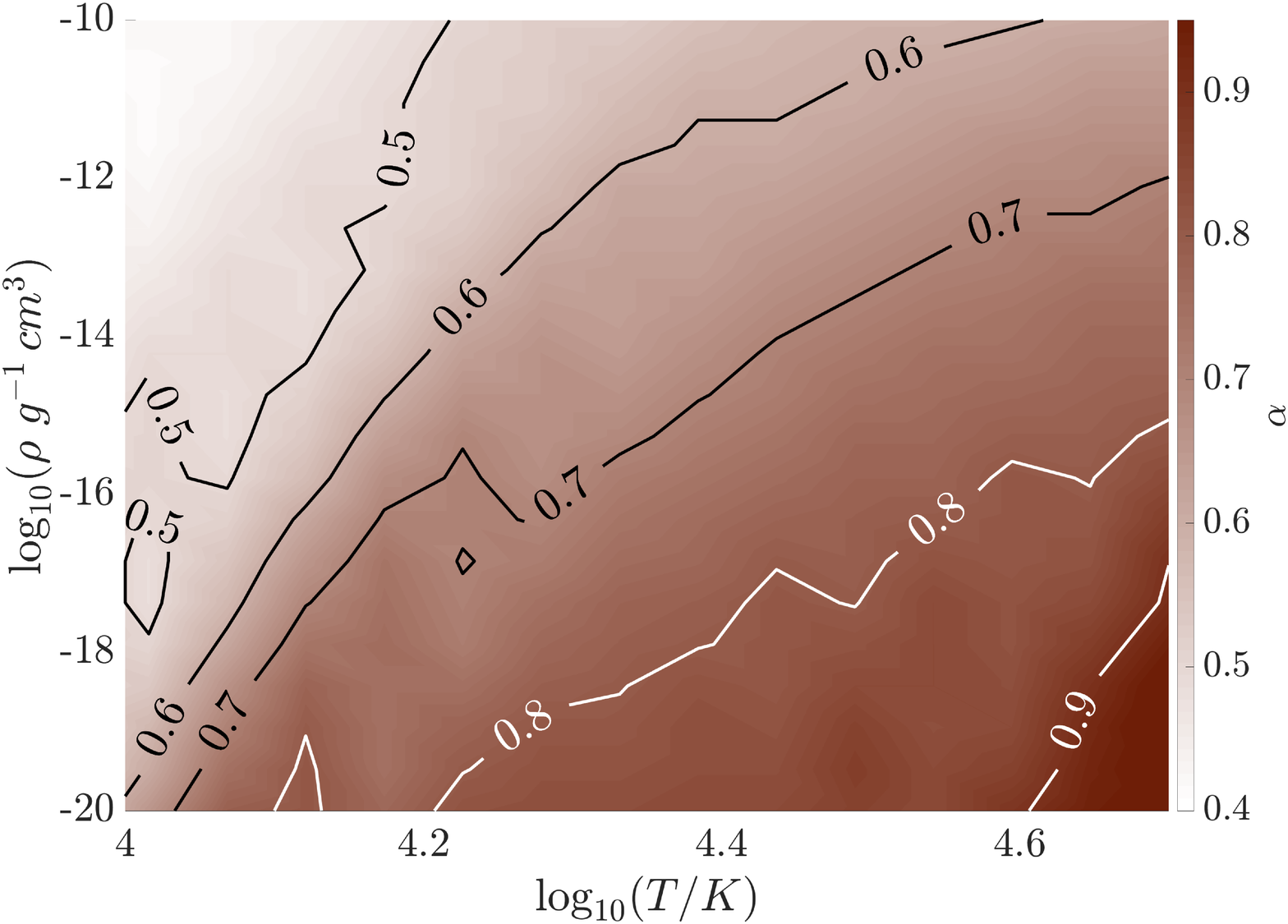}
			\caption{}
			\label{Fig_map_al}
		\end{subfigure}\\%
		\begin{subfigure}[t]{0.5\linewidth}
			\centering
			\includegraphics[width = 0.95\linewidth]{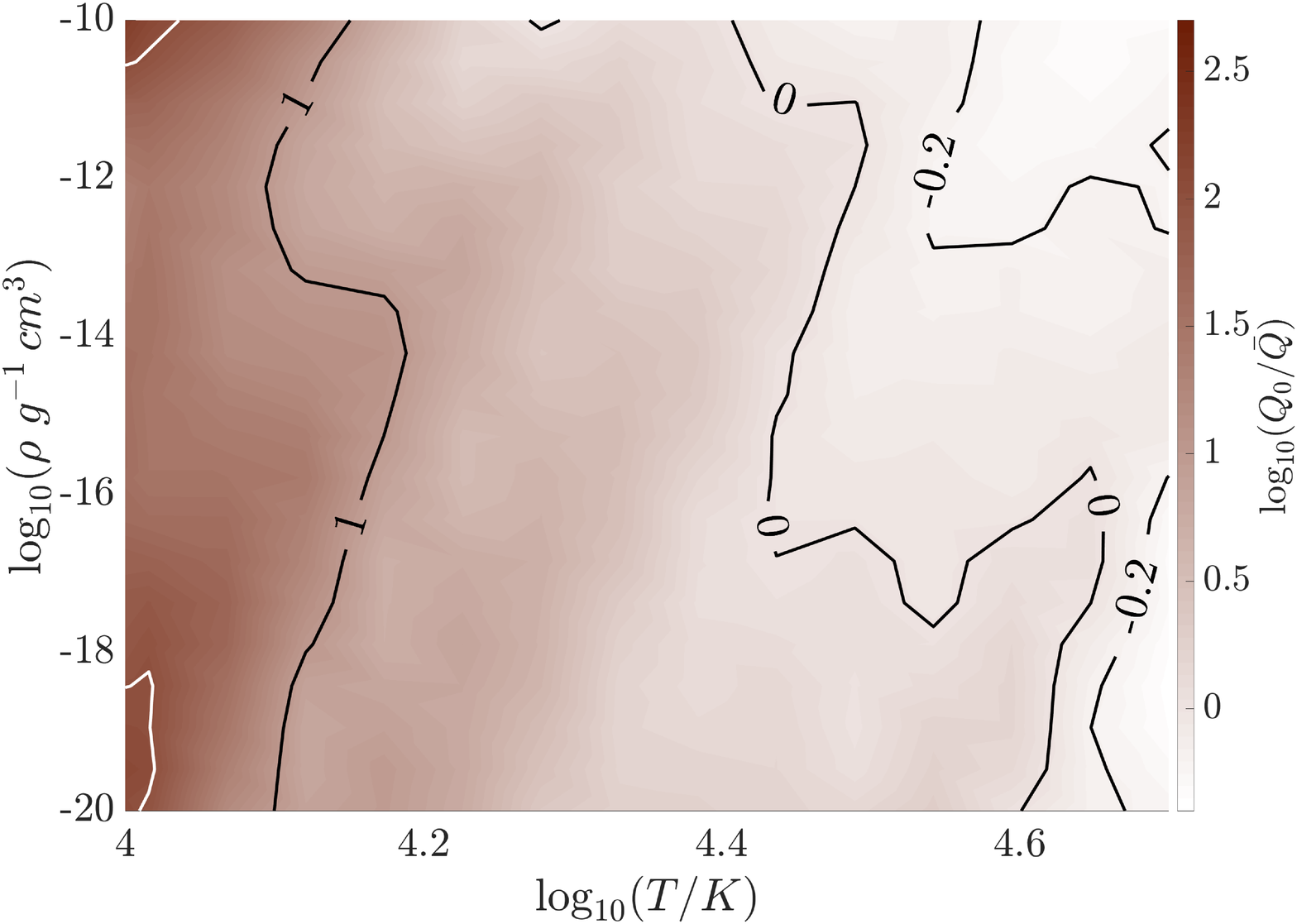}
			\caption{}
			\label{Fig_map_q0}
		\end{subfigure}
	\caption{Colour maps of the line-distribution statistical parameters (a) $\log_{10}\bar{Q}$, (b) $\alpha$, and (c) $\log_{10}(Q_0/\bar{Q})$  computed for a range of temperatures and densities using the Munich line database. Abscissa here measures the logarithm of the temperature in $\UnitT$ and the ordinate gives the logarithm of the density in $\Unitrho$.}
	\label{Fig_map_rest}
\end{figure*}

We apply the method described above to various combinations of temperature and density in the range $\log_{10}(\rho\,[{\rm g^{-1}cm^3}])\in[-20, \,-10]$ and $\log_{10}(T\,[{\rm K}^{-1}])\in[4.0,\; 4.7]$, assuming a solar plasma composition as given by \citet{Asplund_09}.
Using colour maps, we present the results of $\log_{10}\bar{Q}$, $\log_{10}(Q_0/\bar{Q})$, and $\alpha$ in Fig.~\ref{Fig_map_rest}.

Examining these figures, we focus on the region of interest for our initial study of O-type stars in Sect.~\ref{Se_first_appl}. 
Typical surface conditions in O-stars cover the region of $\log_{10}(\rho\,[{\rm g^{-1}cm^3}])\in[-14, \,-10]$ and $\log_{10}(T\,[{\rm K}^{-1}])\in[4.3,\; 4.7]$. 
Inspecting this region in Figs.~\ref{Fig_map_qb}, \ref{sub@Fig_map_al}, we find values of $\alpha \approx 0.5-0.7$ and $\bar{Q} \approx 1000-2000$ that again are in good agreement with previous estimates \citep[e.g.,][]{Gayley_95, Puls_00}. 
Moreover, for the O-star regime, we also find that the Ansatz $Q_0\approx\bar{Q}$ by \citet{Gayley_95} holds reasonably well. 
However, we see that the tabulated values can largely deviate from these often used estimates outside this region.
For example, also in the O-star temperature regime at moderately low densities (typical for the outer wind), we find $\bar{Q}<1000$ and $\alpha> 0.7$. 
As further discussed in Sect.~\ref{Se_first_appl}, such variation of the line-force parameters may then significantly affect the radiation force and alter the wind dynamics. 

\subsection{From tables to opacity}
\label{Se_Table-to-force}

 The tables presented here are now ready to be applied towards computations of the flux-weighted mean opacity, which can then be used in a broad range of applications where the radiation line force needs to be accounted for. In the Sobolev limit, we find a general expression for the flux-weighted opacity diagonal tensor by combining Eqs.~\ref{Eq_kf_kL} and \ref{Eq_Mt_Gayley}:
\begin{equation}
	\label{Eq_opacity_final}
	\bar{\kappa}_{\rm F} = \frac{\mathbf{F}}{F^2}\frac{\kappa_0\bar{Q}}{1-\alpha}\oint\d\Omega \,\mathbf{\hat{n}}\, I^{\rm core}(\mathbf{\hat{n}}) \frac{[1 + Q_0\tau_t(\mathbf{\hat{n}})]^{1-\alpha} - 1}{Q_0\tau_t(\mathbf{\hat{n}})} \, .
\end{equation}
Note that in this general expression, we have allowed $I^{\rm core}(\mathbf{\hat{n}}) = \int\d\nu\,I_\nu^{\rm core}(\mathbf{\hat{n}})$ and $\tau_t(\mathbf{\hat{n}})$ to potentially be functions of the ray direction, whereas the parameters $\alpha$, $\bar{Q}$, and $Q_0$ are all local properties of the gas, which for convenience all have been tabulated using the radial case (Sect.~\ref{Se_line_in_Sobolev}). 
As also noted by \citelc, since the different ray directions effectively serve to provide a scaling factor to the radial streaming case, calibrating our tables using the latter is justified. 

% \dylan{In our numerical calculations, we find appropriate values of $\alpha$, $\bar{Q}$, $Q_0$, and $\tau_{\rm t}$ for each cell based on the local density, temperature, and velocity gradient.}
In our numerical models, for each separate spatial cell at each time-step, we find appropriate values of $\alpha$, $\bar{Q}$, $Q_0$, and $\tau_{\rm t}$ based on the local density, temperature, and line-of-sight velocity gradient.
In practice, values of $\alpha$, $\bar{Q}$, and $Q_0$ are found by interpolations in our pre-described tables (i.e., very similar to how tables of Rosseland mean opacities often are used in various applications).
%In practice, in our numerical calculations, we find the appropriate values of $\alpha$, $\bar{Q}$, and $Q_0$ by interpolations in our pre-described tables (i.e., very similar to how tables of Rosseland mean opacities often are used in various applications). 
In this first study, we further restrict ourselves to analysing the flux-weighted line opacity in the Sobolov limit. 
However, in Sect.~\ref{Se_disc_LDI} we discuss a possible extension also towards non-Sobolev (observer's frame) applications.

\section{First applications}
\label{Se_first_appl}
Generally, self-consistent massive-star wind calculations 
using detailed radiation transfer codes such as \texttt{FASTWIND} \citep{Sundqvist_19}, \texttt{METUJE}  \citep{Krticka_17},  \texttt{PoWR} \citep{Sander_17}, or \texttt{CMFGEN} \citep{Gormaz_21} are highly time-consuming. 
Those computations are also exclusively 1-D and only consider spherically symmetric and steady-state wind outflows. 
Compared to these steady-state models, relaxing a 1-D O-star model by means of the approach introduced here increases the computational speed by a significant factor $> 100$. 
Thus, our new method offers a direct way to straightforwardly take the full line statistics into account also within time-dependent 2-D and 3-D simulations (something which is computationally prohibited when using the time-consuming radiative transfer methods applied in the codes mentioned above).  
Furthermore, in comparison to previous time-dependent line-driven wind models \citep[see, e.g.][]{Feldmeier_17, Dyda_18}, which typically rely on line-force parameters that are pre-described and fixed throughout the simulations (i.e. both in space and time), our new models allow these parameters to vary in space and time, self-consistently computing and updating them according to the local wind conditions.

In this initial study, our key objective is to benchmark our method and tables. 
To achieve this, we investigate the wind outflows of a calibrated O-star model grid used by \citet{Bjorklund_21}.  
The values of stellar luminosity $\L$, stellar mass $\M$, and stellar radius $\R$ of the considered models are summarised in Table~\ref{Tb_O_grid}. 
In addition to studying the behaviour of the line-force parameters, the O-star grid allows us to further validate our technique by comparing mass-loss rates found in our simulations to the values predicted for such stars by other studies using more detailed radiative transfer.  

\begin{table*}[ht!]
\centering
\caption{Resulting mass-loss rates $\dot{M}$ % and terminal wind velocities $\varv_\infty$ 
from our hydrodynamic models for a grid of O-star models with stellar  parameters ($\L$, $\M$, $\R$, and $T_\eff$ ) adopted from \citet{Bjorklund_21}. 
Summarised for comparison are also mass-loss rates $\dot{M}_B$ %and terminal velocities $v_{\infty\, , B}$
predicted by \citet{Bjorklund_21}, and mass-loss rates $\dot{M}_K$ predicted by \citet{Krticka_17}.
Stellar parameters are given in unit of solar luminosity $\L_\odot$, mass $\M_\odot$, and radius $\R_\odot$. }
\begin{tabular}{ c c c c c c c c }
\hline\hline
Model & $\log_{10}(\L/\L_\odot)$ & $\M/\M_\odot$ & $\R/\R_\odot$  & $T_\eff/\UnitT$ &  $\dot{M}/(\UnitMd)$ & $\dot{M}_B/(\UnitMd )$ & $\dot{M}_K/(\UnitMd )$\\ % & $\varv_\infty/ (\UnitV )$ & $v_{\infty\, , B}/(\UnitV )$ \\
\hline
Mod-$0$ & $5.91$ & $58$ & $20$ & $40.0$ & $1.8\cdot 10^{-6}$ & $2.2\cdot 10^{-6}$ & $1.5\cdot 10^{-6}$\\ % & $1600$ & $2300$  \\ 
Mod-$1$ & $4.82$ & $20$ & $8.0$& $32.5$ & $2.2\cdot 10^{-8}$ & $9.1\cdot 10^{-9}$ & $2.4\cdot 10^{-8}$ \\% & $2000$ & $3700$  \\ 
Mod-$2$ & $5.10$ & $27$ & $10$ & $35.5$ & $6.8\cdot 10^{-8}$ & $2.5\cdot 10^{-8}$ & $7.0\cdot 10^{-8}$ \\% & $2200$  & $5300$ \\ 
Mod-$3$ & $5.13$ & $21$ & $14$ & $30.0$ & $7.0\cdot 10^{-8}$ & $4.7\cdot 10^{-8}$ & $8.0\cdot 10^{-8}$ \\ %& $1100$ & $2800$  \\ 
Mod-$4$ & $5.44$ & $31$ & $15$ & $34.6$ & $2.3\cdot 10^{-7}$ & $1.4\cdot 10^{-7}$ & $2.5\cdot 10^{-7}$ \\% & $1700$ & $3000$  \\ 
Mod-$5$ & $5.71$ & $41$ & $15$ & $39.5$ & $8.1\cdot 10^{-7}$ & $1.0\cdot 10^{-6}$ & $6.9\cdot 10^{-7}$ \\ %& $1700$ & $2200$  \\ 
Mod-$6$ & $5.84$ & $58$ & $13$ & $46.0$ & $1.3\cdot 10^{-6}$ & $1.4\cdot 10^{-6}$ & $1.1\cdot 10^{-6}$ \\ %& $1800$ & $3100$  \\ 
\hline
\end{tabular}

\label{Tb_O_grid}
\end{table*}

\subsection{Numerical simulations of line-driven O-star winds}
The radial variation of density and velocity of a spherically symmetric outflow are described by the hydrodynamical equations of mass and momentum conservation:
\begin{equation}\label{Eq_mass_conservation}
\frac{\partial \rho}{\partial t} + \frac{1}{r^2}\frac{\partial}{\partial r}(r^2\rho \varv) = 0\, ,
\end{equation}
\begin{equation}\label{Eq_mome_conservation}
\frac{\partial }{\partial t} (\rho \varv)+ \frac{1}{r^2}\frac{\partial }{\partial r}(r^2 \rho \varv) 
	= -\frac{\partial P}{\partial r} - \rho g + \rho \frac{\kappa_{e}F_{r}}{c} + \rho \frac{\kappa_{{\rm F\,}(r,r)}F_{r}}{c} \, .
\end{equation}
Here, $\varv$ is the radial velocity in $\rm cm\,s^{-1}$, $g = \M G/r^2$ is the gravitational acceleration in $\rm cm\,s^{-2}$, and $P$ is the gas pressure in $\rm g\,cm^{-1}\,s^{-2}$. 
Using the ideal gas law, $P = a^2\rho$, where  $a^2 = k_{B}T/(\mu m_H)$ is the isothermal sound speed, $\mu = 0.66$ is the mean molecular weight of a fully ionised solar plasma, and $m_H$ is the mass of the hydrogen atom. 
We further assume that the winds are isothermal, using the standard assumption (for O-stars) $T  = 0.8T_\eff$ \citep[][]{Puls_00}, where $T_\eff^4 =\L/(4\pi \sigma_{\rm SB} \R^2)$ is the effective temperature of the star.\footnote{In principle, this difference between $T$ and $T_{\rm eff}$ introduces a small inconsistency with the $T = T_{\rm rad}$ assumption that went into the calculation of the tables. However, from test-calculations using also $T = T_\eff$ we have verified that this does not cause any significant differences for the results presented in this section.} 
Finally, $\kappa_\e F_{r}/c$ and $\kappa_{{\rm F\,}r,r}F_{r}/c$ are the two components of radiation acceleration from Thompson scattering and spectral lines, respectively, with $F_{r} = \L/(4\pi r^2)$ the radial radiation flux. 

Assuming a uniformly bright star with $I^{\rm core} = F/\pi$ for $\mu\in[\mu_*,\,1]$, where $\mu_*^2  = 1 - \R^2/r^2$, Eq.~\ref{Eq_opacity_final} gives:
\begin{equation}
	\label{Eq_kappa_in_hydro}
	\kappa_{{\rm F\,}(r,r)} =  \kappa_0\, \eta\, \frac{\bar{Q}}{1 - \alpha} \frac{(1 + Q_0\,\tau_t)^{1-\alpha}  - 1}{Q_0\,\tau_t}~ \Unitkap\, ,
\end{equation}
where the Sobolev optical depth parameter from Eq.~\ref{Eq_dimless_t} reduces to $\tau_t = c\,\rho\kappa_0 / |{\d \varv}/{\d r}|$. 
Here $\eta$ is a geometrical correction factor accounting for the finite size of the star, the exact computation of which requires numerical evaluation of the solid angle integral given in Eq.~\ref{Eq_opacity_final}. 
However, assuming $Q_0\tau_t\gg1$ when evaluating $\eta$ provides an analytic estimate (\citealt{Pauldrach_86}; see also \citealt{Cranmer_96}, section 2.2.4):
\begin{equation}
\label{Eq_finite_disk_limit}
\eta(r) = \frac{(1+\sigma)^{1+\alpha} - (1+\sigma\mu_*^2)^{1+\alpha}}{(1+\alpha)(1-\mu_*^2)(1+\sigma)^\alpha\sigma}\, ,
\end{equation}
with homologeous wind expansion parameter
\[\sigma = \frac{\d \ln \varv}{\d \ln r} -1\, .\]

For simplicity, we introduce the effective gravity $g_\eff(r) = \M G [1 - \Gamma_\e(r)]/r^2$, where $\Gamma_\e(r) =  {\kappa_\e(r)\L}/{(4\pi \M G c)}$ is the Eddington ratio, so that Eq.~\ref{Eq_mome_conservation} reads as:
\begin{equation}\label{Eq_mome_conservation_eff}
\frac{\partial }{\partial t} (\rho \varv)+ \frac{1}{r^2}\frac{\partial }{\partial r}(r^2 \rho \varv) 
	= -\frac{\partial P}{\partial r} -\rho g_\eff+ \rho \frac{\kappa_{{\rm F\,}r,r}F_{r}}{c}\, .
\end{equation}
To compute the effective gravity, we use the appropriate pre-tabulated values of the Thompson scattering opacity $\kappa_\e$.

We use the \texttt{midpoint} time integration, with \texttt{TVDLF} Riemann solver, and  a \texttt{VANLEER} flux limiter \citep{vanLeer_79}.
As described in the previous section, computation of $\kappa_{{\rm F\,}(r,r)}$ requires knowledge of the local velocity gradient.
We compute this by applying a simple finite-differencing for non-constant intervals
%\footnote{In our numerical setup, we apply a constant stretching factor $\Delta r_{i+1}/\Delta r_{i} = 1.002$ between subsequent cells $i,\,i+1$ to resolve the transonic region.}  
to the velocity field at each numerical time step \citep{Sundqvist_70}.
We also limit the integration time step $\d t$ to avoid information propagating over multiple cells in a single step by:
\[
\d t = \min \left(\d t_{\rm CFL},\, {\rm CFL}\times\sqrt{\frac{c \Delta r}{\kappa_{{\rm F\,}(r,r)}F_{r}}}\right)\, ,
\]
where $\d t_{\rm CFL}$ is the time step set by the standard Courant--Friedrichs--Lewy (CFL) condition \citep{Courant_28}, using here ${\rm CFL} = 0.3$, and $\Delta r $ is the width of one numerical cell. 

Each simulation covers a radial range extending from a lower boundary at $\R$ to an outer boundary at $6\, \R$. For the restricted parameter range investigated in this paper (essentially O-stars), such an outer boundary is sufficient to ensure that the winds have reached velocities sufficiently close to their asymptotic values (i.e. they are not significantly accelerating at the outer boundary). In our numerical setup, we further apply a constant stretching factor $\Delta r_{i+1}/\Delta r_{i} = 1.002$ between subsequent cells $i,\,i+1$ to resolve the transonic region.

\subsection{Initial and boundary conditions}

%\textit{Initial conditions:}
We set the initial velocity using the so-called $\beta$-law:
\begin{equation}\label{Eq_initial_velocity}
	\varv_\beta(r) = \varv_\infty\left(1 - b\frac{\R}{r}\right)^\beta\, ,
\end{equation}
with $\beta= 0.5$ and terminal velocity $\varv_\infty^2 = 2\R\, g_\eff(\R)$.
To avoid infinite density in the first computational cell, we chose $b$ such that $\varv_\beta(\R) = 10\UnitV$. The initial density follows
\begin{equation}\label{Eq_initial_density}
\rho(r)  = \rho_b\frac{\varv_\beta(\R)}{\varv_\beta(r)}\left(\frac{\R}{r}\right)^2\, , %\frac{\dot{M}}{4\pi r^2 \varv_\beta(r)}\, 
\end{equation}
where $\rho_{b}$ is an input parameter. 
In principle, we could use an analytic solution of mass loss (see \citecak\ for more details) for the initial condition, rather than setting the input value $\rho_b$.
However, we choose not to do so to investigate the relaxation of the time-dependent solutions starting from different initial structures. 
Indeed, starting the same setup with different values of $\rho_b$ (varying it by an order of magnitude), we find almost identical final structures, thus confirming that our results are quite independent of the specific initial conditions.

%\textit{Boundary conditions:} 
We set the upper boundary conditions by simple extrapolations of velocity and density \citep[see, e.g.][]{Poniatowski_21}.
In contrast to our previous simulations, we here allow the lower boundary conditions to self-adjust to the overlaying wind by updating the density of the first ghost cell $\rho_1$ when required. 
This is done if the following condition is met: $| \rho_1/(2\bar{\rho}_s) - 1| \geq 20\%$, where $\bar{\rho}_s$ is the average density at the sonic point $\varv(R_s)^2 = a^2$, with the averages taken over a dynamical time-scale 
$ t_{\rm dyn} = \R/\varv_\infty$. 
Whenever this condition is met, we set $\rho_1 = 2 \bar{\rho}_s$. 
The remaining ghost cells at the lower boundary are then set by assuming quasi-hydrostatic equilibrium for the density, whereafter the velocity is determined from mass conservation. In comparison to our previous models with $\rho_1$ fixed throughout the simulation duration, this procedure allows the wind to adjust itself somewhat better to initial conditions that may be rather far away from the final relaxed state. 
%the density in the remaining cell $i$ are simply set by the exponential stratification, with 
%then set by density scale-height $H = 2 a^2/ [\R\, g_\eff(\R) ]$
%
%\[\rho_i = %\rho_1\e^{-\sfrac{\R}{H}\left(1 - %\sfrac{\R}{r_i}\right)} \, ,\]
%
%and the velocity in these cells is then simply set by the mass continuity equation.}

\subsection{Mod-$0$}
\label{Se_Model-0}
Following the above setup, we first present a numerical model (Mod-0) with stellar parameters roughly corresponding to $\zeta$ Puppis, which is often referred to as a "prototypical" O-star.
The distinctive point of our simulation, as mentioned above, is that here the line-distribution parameters are interpolated\footnote{We here use linear interpolation in logarithmic space} from tables and thus consistent with the local conditions in the wind; therefore, they vary in time and space. That is, as outlined in Sect.~\ref{Se_Flux_mean_opacity}, we are computing $\alpha(\rho,T)$, $\bar{Q}(\rho,T)$, $Q_0(\rho,T)$ at each radius and each time-step in the simulation, and then we use Eq.~\ref{Eq_kappa_in_hydro} to obtain the corresponding $\kappa_{\rm F}(r,t)$ that is consistent with the local hydrodynamical structure.
Since the local values of $\alpha$, $\bar{Q}$, and $Q_0$ set the scale of the line force, it is natural that the radial and temporal variation of these parameters then also reflects upon the self-consistent mass loss and velocities of the models. 
\begin{figure*}%[ht!]
	\centering
	\includegraphics[width=\textwidth]{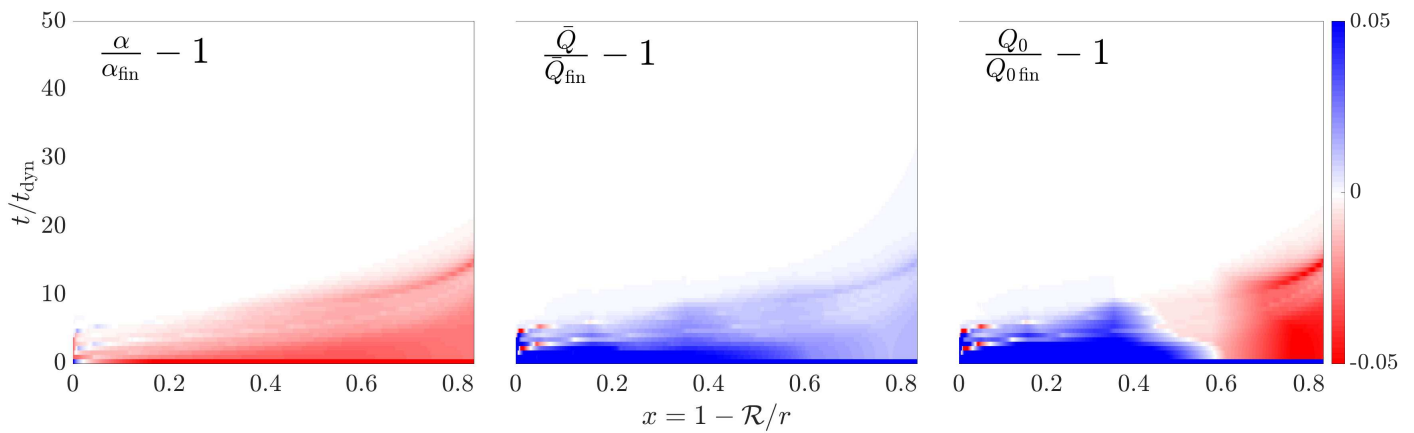}
	\caption{Colour map showing the radial and temporal variation of the line distribution parameters. Left to right we show $\alpha$, $\bar{Q}$, and $Q_0$ radially normalised to their respective final profiles $\alpha_{\rm fin}$, $\bar{Q}_{\rm fin}$, and $Q_{0\,\rm fin}$. }
	\label{Fig_param_var_map}
\end{figure*}
To illustrate this variation, (in Fig.~\ref{Fig_param_var_map}) we plot colour maps of all three parameters as a function of scaled radius $x = 1 - \R/r$ and simulation time $t$ in units of dynamical time-scale $t_{\rm dyn}$. For better visualisation, we normalised each parameter to their radial profile from the final snapshot. As seen in the figure, after adjusting to the new conditions, all three parameters relax to a quite steady behaviour in time.  
To ensure that our models are well relaxed, we typically run models for at least $90\,t_{\rm dyn}$, although in the O-star regime, the relaxation is typically faster and a (quasi) steady-state `final' configuration is reached much earlier than this.
Because the relaxed radial profiles of $\alpha$, $\bar{Q}$, and $Q_0$ are significantly different from the typically assumed constant profiles, let us next discuss how this variation influences the wind dynamics. 

\begin{figure}%[ht!]
	\centering
    \begin{subfigure}[t]{0.5\textwidth}
	    \centering
	    \includegraphics[width=\linewidth]{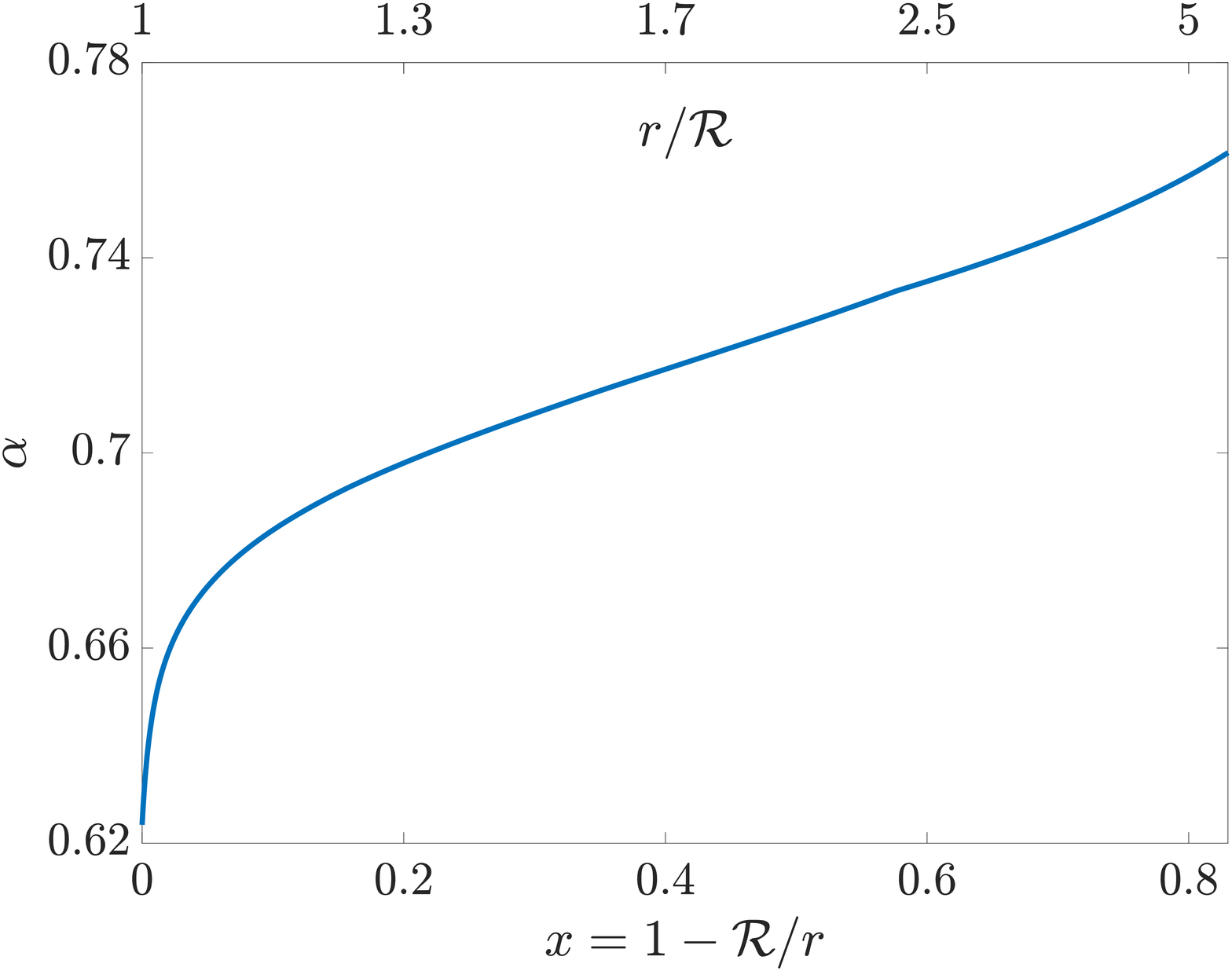}
	    \caption{}
	    \label{Fig_alpha_fin}
    \end{subfigure}\\ \vspace{8pt}
    \begin{subfigure}[t]{0.5\textwidth}
	    \centering
	    \includegraphics[width=\linewidth]{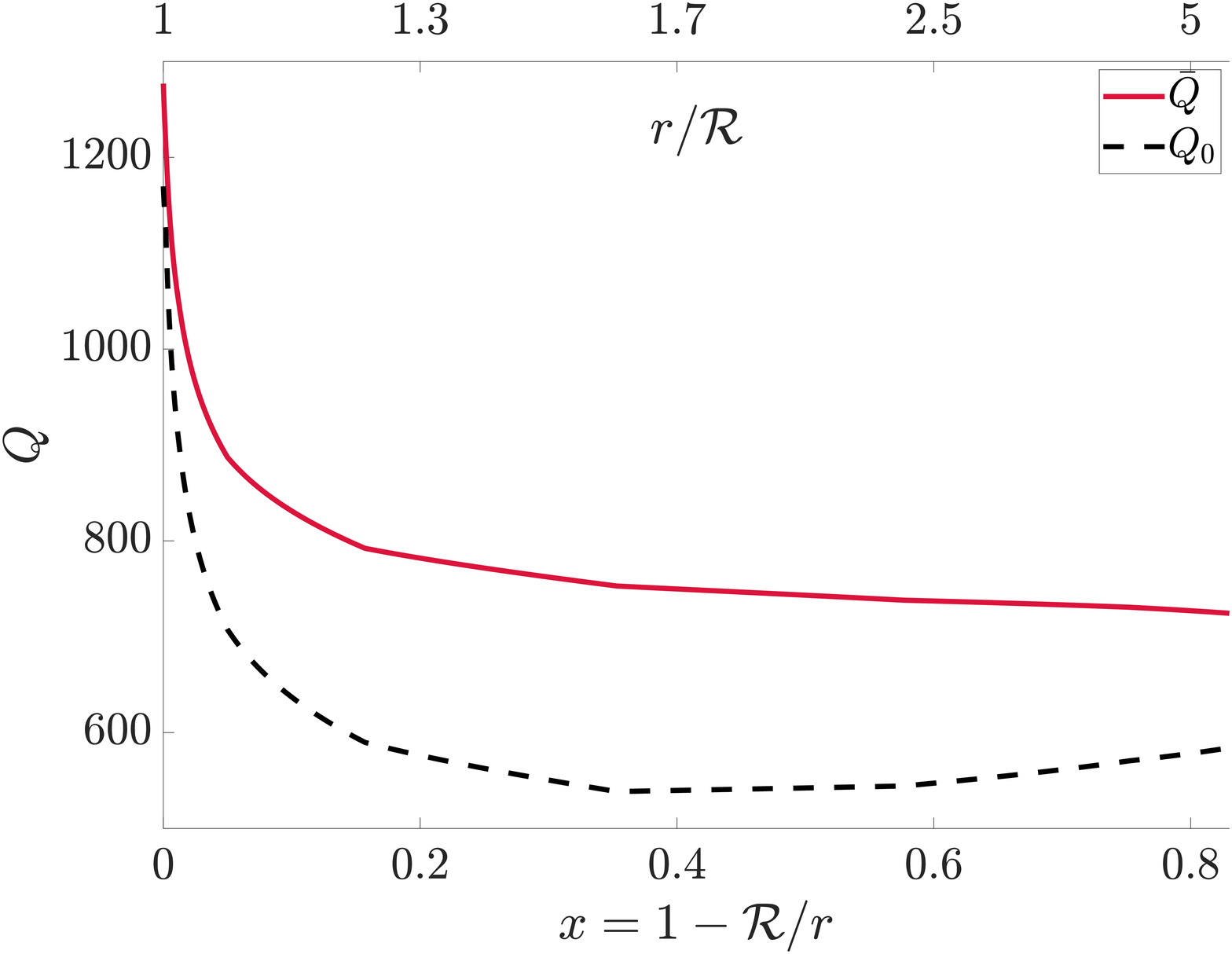}
	    \caption{}
	    \label{Fig_qq_fin}
    \end{subfigure}
    \caption{Radial profile of the line-distribution parameters of the final snapshot of Mod-$0$. Plot (a) shows the $\alpha$ parameter, (b) gives the $\bar{Q}$ (solid-red line) and $Q_0$ (dashed-black line) parameters.}
\end{figure}

We show the radial profile of  $\alpha$ in Fig.~\ref{Fig_alpha_fin}.
Near the lower boundary, we observe that $\alpha$ starts with values reasonably close to the typically expected $2/3$ \citep[see, e.g.,][]{Puls_08}; however, when the radius increases, so does $\alpha$. 
A possible interpretation of this rise comes from the distribution of optically thick and thin lines that contribute to the line force. 
Splitting Eq.~\ref{Eq_Mt_sum} into two parts 
\begin{equation}\label{Eq_Mt_split}
M(\tau_t) \approx \sum_{i}^{q_i \tau_t \ll 1} w_{\nu_i} q_i 
+ \sum_{j}^{q_j \tau_t \gg 1} \frac{w_{\nu_j}}{\tau_t}\;,  
\end{equation}
we notice how the optically thin part has no direct 
dependence on the velocity gradient ($\alpha=0$), whereas the optically thick part is directly proportional to it ($\alpha=1$). Following \citet{Puls_00}, we may thus interpret the value of $\alpha$ as the fraction of the line force that comes from optically thick lines to the total line force.  
As such, we can associate the increase of $\alpha$ with radius as a change of the line-force character from having relatively even contributions from weak and strong lines to being dominated by more optically thick lines.
Indeed, driving the outer wind of O-stars by only a few strong lines is consistent with previous models focusing exclusively on stationary winds, for example, see Fig. 9 in \citet{Krticka_06} and the general discussion in Sect. 2 of \citet{Puls_08}\footnotemark.
Moreover, this general interpretation seems to be supported by observations of O-stars, where we see only a few strong spectral lines formed at a high wind velocity. 
Also, since differences in abundances/ionisation stages can significantly influence these relatively few outer-wind driving lines, a significant scatter of terminal wind speeds can be expected \citep[see also][]{Puls_00}, as is also observed \citep{Garcia_14}.
\footnotetext{Although we also note that an earlier study by \citet{Schaerer_94} does seem to rather imply a radial decline of $\alpha$ within a stationary wind; the way of practically deriving the actual values for $\alpha$ in that study is, however, quite different from here, and as such it is difficult to assess where this difference in results might come from.}

Inspecting Fig.~\ref{Fig_qq_fin}, we further find a decrease of both $\bar{Q}$ and $Q_0$ towards the outer wind. 
Generally, the decrease of $\bar{Q}$ suggests either that the contributing lines in the outer wind are weaker as compared to the inner wind or that fewer lines are contributing to the total line force.
Both an increase of $\alpha$ and a decrease of $\bar{Q}$ lead to a lower radiation force, which in turn also reduce the wind velocity. 
However, a lower $Q_0$ can counteract this, as this leads to the opposite effect, causing an increase in the radiation force (by effectively reducing the line optical depth). 

In principle, as each of the parameters depends on the local wind conditions while also affecting these local conditions, they form feedback loops with the wind structure. 
To understand this, let us consider the effects of varying each parameter separately in Eq.~\ref{Eq_kappa_in_hydro}.
Then an increase in $\alpha$ leads to a decrease in the line opacity, which affects the velocity and density profile. 
However, in the O-star wind regime, an increase in density reduces $\alpha$ (see Fig.~\ref{Fig_map_al}), thus forming positive feedback. Similarly, from Fig.~\ref{Fig_map_qb} we see that increasing density also increases $\bar{Q}$, and thus the line opacity, again forming a positive feedback loop. 
By contrast, $Q_0$ and the line force form a negative feedback loop, where increasing $Q_0$ rather decreases the line opacity. 
A self-regulatory system forms through the interaction of these positive and negative feedback loops, which ultimately here determines the final mass-loss rate and velocity law of the wind.

\begin{figure}
    \centering
	\begin{subfigure}[t]{0.5\textwidth}
		\includegraphics[width=\linewidth]{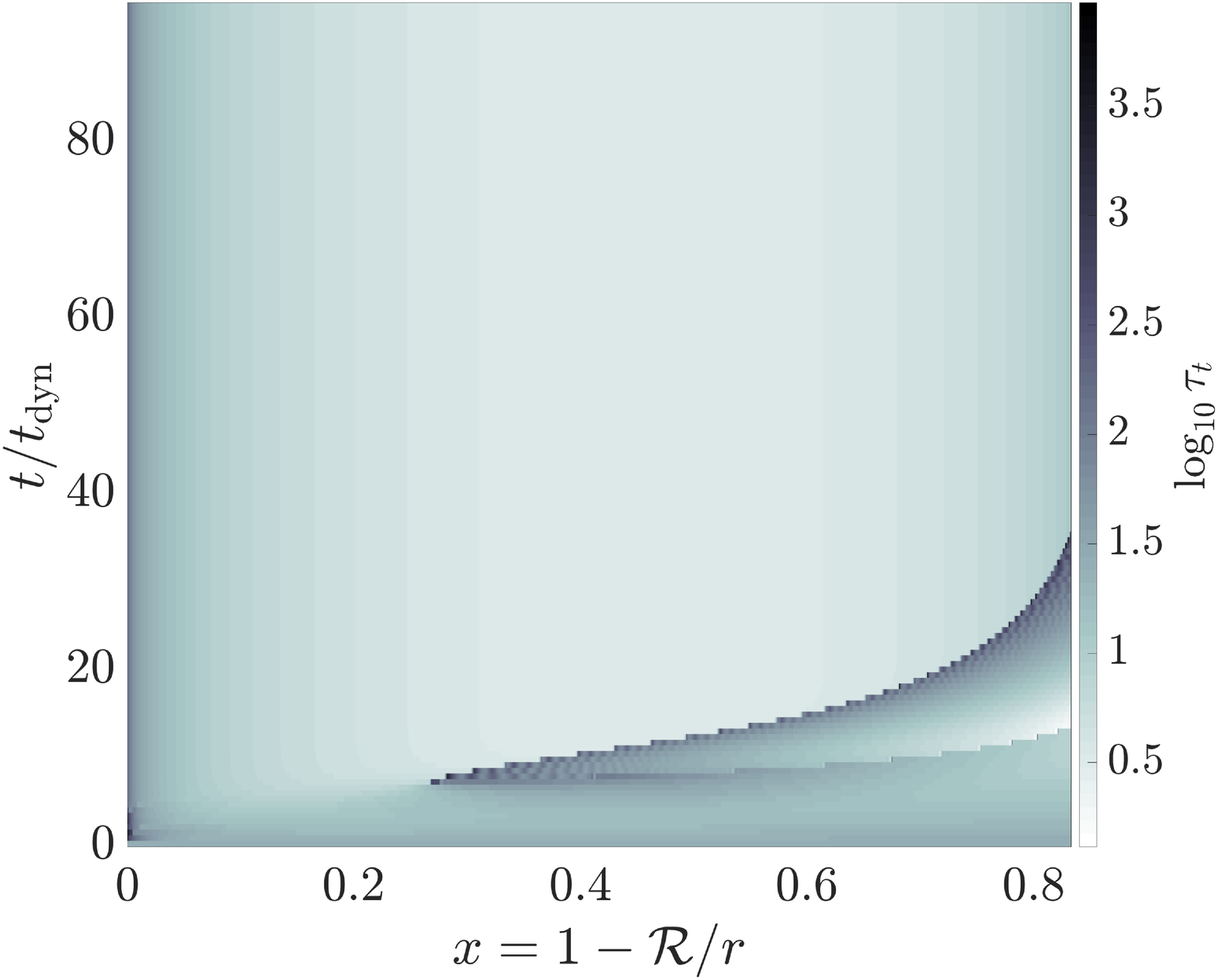}
		\caption{}
		\label{Fig_t_map}
	\end{subfigure}\\ \vspace{8pt}
	\begin{subfigure}[t]{0.5\textwidth}
		\includegraphics[width=\linewidth]{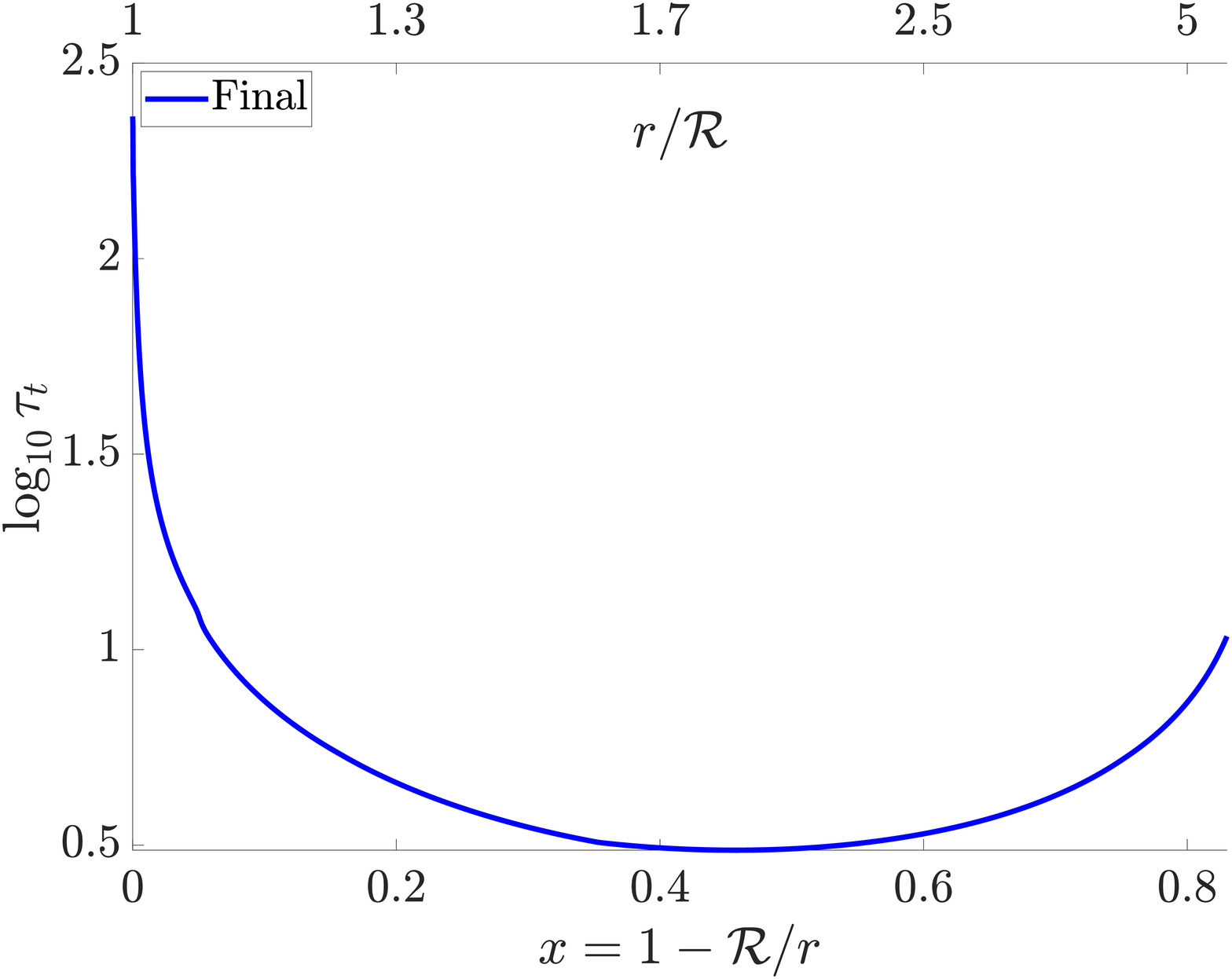}
		\caption{}
		\label{Fig_t_fin}
	\end{subfigure}
    \caption{ Colour map showing the radial and temporal variation of optical depth parameter $\tau_t$ (a). Extrema found in this map are $\max(\log_{10}\tau_t) = 3.95$ and $\min(\log_{10}\tau_t) = 0.12$.
	Radial profile of $\tau_t$ from the final snapshot (b).}
    \label{Fig_ts}
\end{figure}

In Fig.~\ref{Fig_t_map}, we show the relaxation of the optical depth parameter $\tau_t$. 
Similar to the relaxation of the line-distribution parameters (Fig.~\ref{Fig_param_var_map}), we observe a transient behaviour from the initial condition, which takes about $40\,t_{\rm dyn}$ to fully vanish. After this, a near-equilibrium configuration is reached. 
Importantly, the value of $\tau_t$ during the simulation never exceeds the range initially used to compute the line-distribution parameters.
As such, our tabulations always cover the correct range in $\tau_t$ (as well as in density, velocity gradient and temperature).\footnote{This is reassuring for the benchmarking purposes of this section, although (as already mentioned in the previous section), due to the nature of the line-force as a function of $\tau_t$, even application of the tables outside initially assumed range of $\tau_t$ should return reliable results.}
Inspecting the final, relaxed values of $\tau_t$ in Fig.~\ref{Fig_t_fin}, we further note that the final saturation of $M(\tau_t)$ at $\tau_t\ll\tau_{t\,{\rm trans}}$ is not reached for this particular simulation. Indeed, the minimum values still have $Q_0 \tau_t \gg 1$, and as such the full O-star wind resides in the power-law portion of $M(\tau_t)$, as also previously noted, for instance by \cite{Gormaz_19}.
However, we emphasise that this is not guaranteed always to be the case, especially not for more complicated situations where the outflow might not be stationary and smooth, but rather time-dependent and clumpy with co-existing regions of (much) higher and lower densities. In such cases, very low values of $\tau_t$ would systematically be found in the low-density parts, and $M(\tau_t)$ might there readily reach the regime where it saturates to a maximum value $\bar{Q}$.

\begin{figure}%[ht!]
\centering
    \begin{subfigure}[t]{0.5\textwidth}
	%    	\centering
		\flushleft
	    \includegraphics[width=\linewidth]{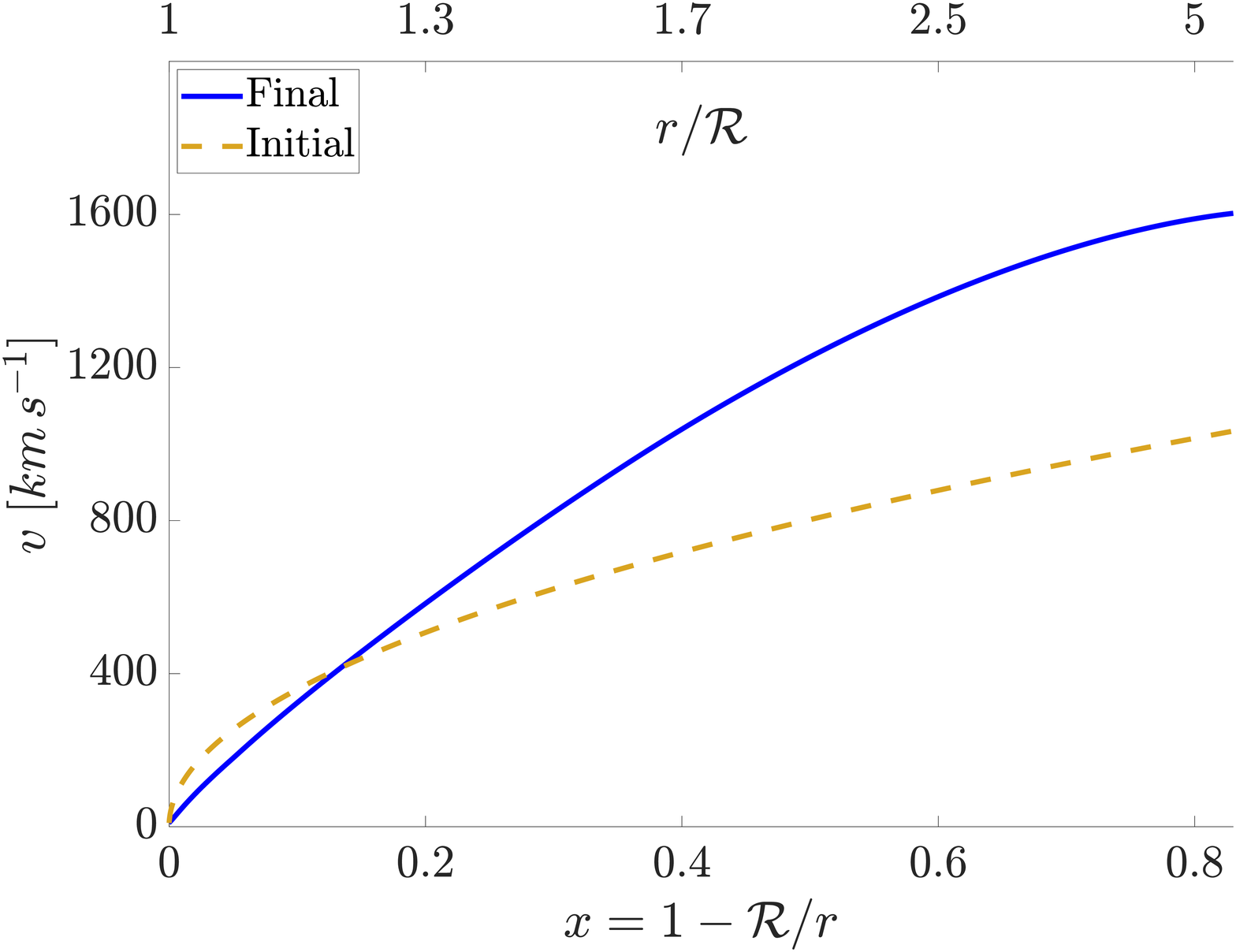}
	    \caption{}
	    \label{Fig_ve_relax}
    \end{subfigure}\\\vspace{8pt}
	\begin{subfigure}[t]{0.5\textwidth}
		\centering
	    \includegraphics[width=\linewidth]{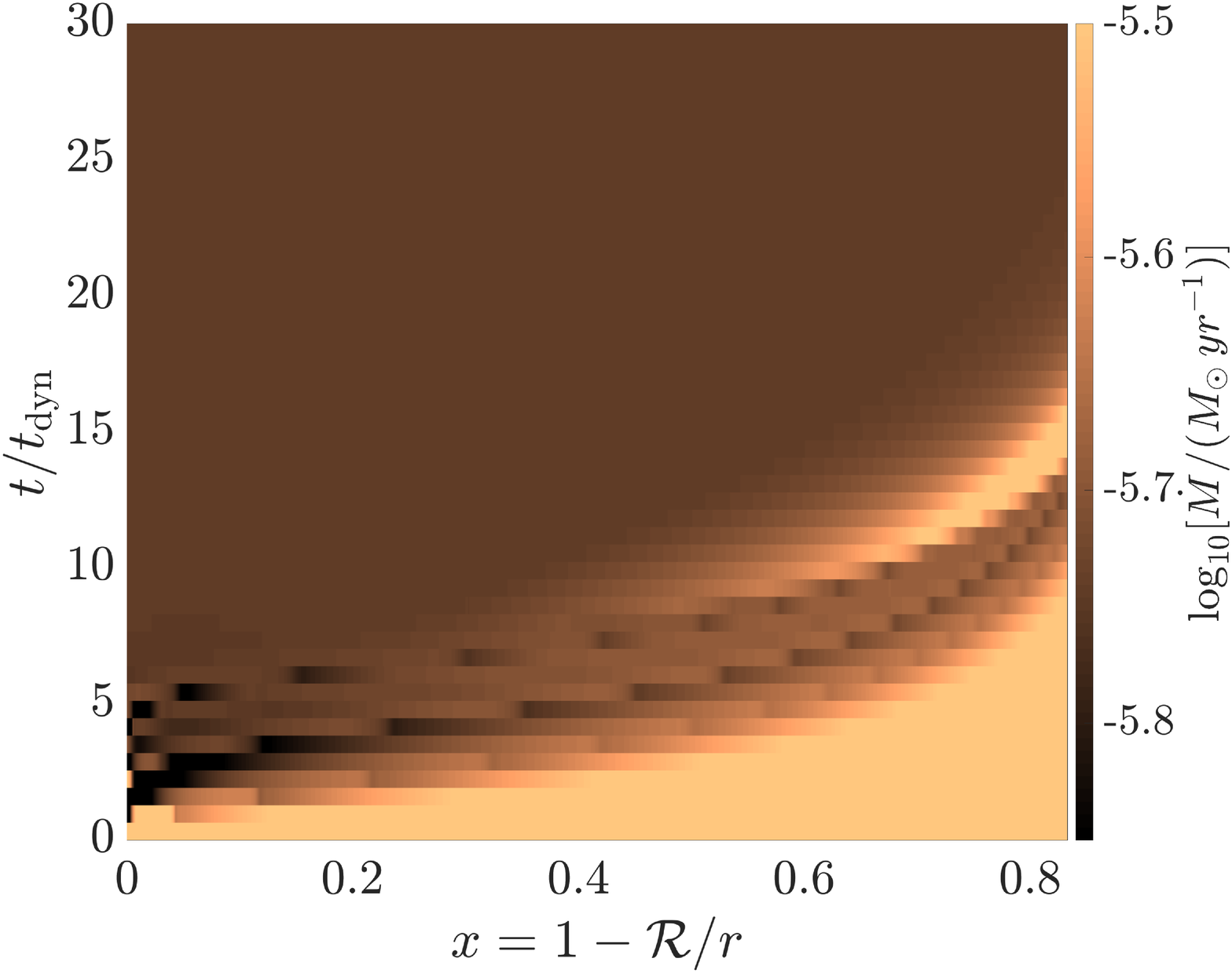}
	    \caption{}
	    \label{Fig_Md_relax}
    \end{subfigure}\\ 

\caption{(a) Radial velocity profiles of the initial conditions in bashed-orange line and the final snapshot in solid-blue line. (b) Colour map showing the relaxation or the radial profile of the logarithm of the mass-loss rate spanning the range from the initial conditions up to $30\,t_{\rm dyn}$.}
\end{figure}

The resulting velocity structure can be seen in Fig.~\ref{Fig_ve_relax} (solid blue line), where we show the velocity in $\UnitV$ as a function of $x$, again for the final snapshot.
For comparison, we show the initial velocity (dashed golden line) prescribed by Eq.~\ref{Eq_initial_velocity}. 
As is evident from the comparison, the final velocity profile does not resemble the initially assumed one very much, exhibiting a different acceleration profile and a higher terminal velocity of $\varv_\infty \approx 1600\UnitV$.
An increase in terminal velocity between the initial (\citecak) and the final profile is typically expected because of the finite disk factor $\eta$ \citep{Pauldrach_86, Friend_86}.
However, variation of $\alpha$, $\bar{Q}$, and $Q_{0}$ will also affect the terminal velocity.
To differentiate between effects introduced by the finite disc factor from the effects introduced by variation of line distribution parameters, we also compute a `reference' model, which uses the same stellar parameters as the Mod-$0$, but where we keep line-distribution parameters constant throughout the simulation. 
In this reference model, we use the near lower boundary conditions of the original model to set the values $\alpha = 0.66$, $Q_0 = \bar{Q} = 1100$.
This then results in a terminal velocity of $\varv^{\rm ref}_\infty \approx 3000\UnitV$. Clearly, compared to the reference model, the terminal velocity is reduced by a factor of $\sim2$. Since the only difference between the two models is how we treat the line-distribution parameters, this stems directly from their specific radial behaviour. 
% Comparing the resulting terminal velocity to the value predicted by \citet{Bjorklund_21} for the same model (see  Table~\ref{Tb_O_grid}), we find a decrease of 

Further, to demonstrate that the wind velocity and density also relax as line-distribution parameters reach their final configuration, in Fig.~\ref{Fig_Md_relax} we plot a colour map of radial and temporal variation of the mass-loss rate zooming in on the first $30\,t_{\rm dyn}$.
The axes in this figure are the same as in Fig.~\ref{Fig_param_var_map}.
Like for the line-force parameters the relaxation takes around $20\,t_{\rm dyn}$ 
%  initially $\dot{M} = 4.7\cdot 10^{-6}$ to 
after which the model relaxes to a final self-consistent mass loss of $\dot{M} = 1.8\cdot 10^{-6}\UnitMd$.
This mass loss then stays constant and only differs by about $20\%$ from the value predicted by \citet{Bjorklund_21} and \citet{Krticka_17} for the same luminosity (see Table~\ref{Tb_O_grid}).
Such an overall good agreement of our models with previous mass-loss results re-emphasises the general quality of the method. Indeed, as now shown, our computed mass-loss rates agree quite well with the predictions by \citet{Bjorklund_21} and \citet{Krticka_17} in the complete O-star domain.

\subsection{Grid of O stars}\label{Se_model_grid}

Using the same setup as for Mod-$0$, we compute numerical models for the stellar parameters presented in Table~\ref{Tb_O_grid}. 
In the previous section, we considered in detail a specific model, pointing out that the discussed behaviours are common for all models within our grid. 
Specifically, in all models, $\alpha$ increases outwards in the wind, whereas both $\bar{Q}$ and $Q_0$ decrease. Near the lower boundary, we find values of all three line-force parameters within their previously estimated range.
However, starting just slightly above the stellar surface, they begin to deviate from this range by significant amounts.
Across all O-star models, we find $Q_0 \approx \bar{Q}$ only in the near-surface region, whereas throughout the rest of the wind $Q_0 < \bar{Q}$.  
In general, this means that studies using the standard Ansatz $Q_0 = \bar{Q}$ tend to overestimate the effective Sobolev optical depth $Q_0\tau_t$ of the line ensemble in low-density regions, which can then lead to an inconsistent radial line force.

Furthermore, although the typical conviction is that NLTE effects are essential in setting the dynamics of O-star winds, our model-grid analysis seems to suggest the opposite, namely that an LTE-based line force (explicit assumption in Sect.~\ref{Se_discrete_sum}) sufficiently well captures the fundamental dynamics of (at least) the near-star region, which essentially sets the mass-loss rate of the model. This suggestion stems from the (on average) tight agreement of the mass-loss rates found in our models with the predictions derived from the full NLTE models by \citet{Krticka_17} and \citet{Bjorklund_21}.

\begin{figure}[ht!]
\centering
\includegraphics[width=0.95\linewidth]{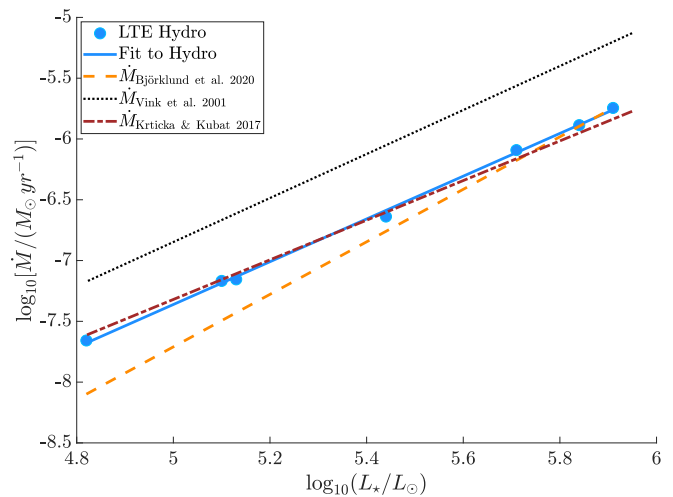}
\caption{ Comparison of the mass-loss rates form our grid of models (blue dots) to the mass-loss the values predicted in earlier works by \citet{Bjorklund_21}, \citet{Krticka_17}, and \citet{Vink_01}. We plot the mass loss-luminosity relations predicted by these authors in orange dashed, maroon dotted-dashed, and black dotted lines, respectively.
We also over-plot the linear fit to our results (solid blue line).}
\label{Fig_Mdot_grid}
\end{figure}

To illustrate this point, we plot in Fig.~\ref{Fig_Mdot_grid}  the resulting mass-loss rates as a function of luminosity.
In addition, we performed a linear fit plotted as a blue line
% % 
% \begin{equation}
% \label{Eq_mass-luminocity}
% \log_{10}\left(\frac{\dot{M}}{1\M_\odot\;yr^{-1}}\right) =  -(5.6\pm 0.05)  + (1.76\pm 0.075)\log_{10}\left(\frac{\L}{10^{6}\L_\odot}\right)
% \end{equation}
and compare this mass loss-luminosity relation to the similar relations by \citet{Krticka_17} in dashed-maroon line and by \citet{Bjorklund_21} in dashed-orange line.   
We indeed observe exceptionally intimate agreement between our and the \citet{Krticka_17} relations and also a close correlation with the \citet{Bjorklund_21} results, with better agreement at the high luminosity end (see also below).
For comparison, we also plot the mass loss-luminosity relation by \citet{Vink_01} in the dotted-black line. 
As compared to \citet{Vink_01}, our models have a systematically lower mass loss by about a factor of approximately $3$.
Such a `factor of 3' difference in the O-star regime is (nowadays) quite well-established \citep[see, e.g.][]{Najarro_11, Sundqvist_11, Surlan_13, Cohen_14, Hawcroft_21}, and our results thus suggest that this reduction is not a consequence of NLTE or co-moving frame transfer effects. 
Rather our models seem to suggest that it is of prime importance to require local dynamical consistency \citep[as in the models by][and here]{Krticka_17, Bjorklund_21}, rather than relying on a global energy constraint \citep[as in the models by][]{Vink_01}, when computing the mass-loss rates (for O-stars). 
We also note that for lower luminosities, our rates start to become higher than those predicted by \citet{Bjorklund_21}, which, most likely, is a direct result of the characteristic radiation-force dip in near-sonic regions observed in non-Sobolev models \citep{Bjorklund_21}.
Interestingly, our results thus suggest that this effect is more significant for lower luminosities and mass-loss rates, as already pointed out by \citet{Owocki_99}. 
Furthermore, \citet{Krticka_17} scale their computed non-Sobolev line force to the corresponding Sobolev one, which shifts their critical point (where the mass loss is set) upstream of the sonic point. 
As a result, their models might be considered as `effectively Sobolev-based' and the nature of their steady-state solutions may thus be quite similar to those presented here, consequently giving a possible explanation to why we obtain such a tight agreement with these authors over the complete O-star domain. Regarding terminal wind speeds, we note that we generally obtain somewhat lower values than those by \citet{Bjorklund_21}. This may stem from differences in the modeling techniques (e.g., CMF transfer or our LTE assumption, for the latter see discussion in Sect.~\ref{Se_dinc_limitations}), and should be further investigated in future work.  

In summary, the overall agreement with these previous works in the well-studied O-star domain provides good support for the general method developed here. 
We remind the reader, however, that our prime goal in this section has not been to provide new velocities and mass-loss rates for O-stars, but rather to test our new (fast and efficient) method against models based on more detailed radiative transfer (but which are limited to 1-D steady-state systems). The key point of the analysis in this section is thus that, with the above benchmarks now in hand, our method is ready for exploitation also in more complicated situations requiring multi-D RHD setups \citep[e.g.,][]{Moens_22}.
%(e.g., Moens, Poniatowski, et al., in prep.). 

\section{Discussion}
\label{Se_discussion}

\subsection{Comparison to \citeauthor{Lattimer_21}}
\label{Se_disc_comp_to_lc}

Using a similar set of assumptions, \citelc\ recently computed force multipliers from an atomic database with approximately the same number of spectral lines as ours. 
However, a key distinction between theirs and our work is that they fit the force multipliers using an ad-hoc functional form that seems to have no apparent relation to an underlying line-distribution function (their eq. 37).
By contrast, the functional form we fit here (Eq.~\ref{Eq_Mt_Gayley}) can be directly derived from assuming a distribution of lines according to Eq.~\ref{Eq_dN}. 
As such, our study becomes much simpler to generalise both toward general multi-D radiation-hydrodynamics simulations (see discussion in Sect.~\ref{Se_first_appl}) and toward situations wherein a non-Sobolev line-force must be considered (see discussion in Sect.~\ref{Se_disc_LDI}).

\begin{figure}[ht!]
	\centering
	\begin{subfigure}[t]{\linewidth}
		\includegraphics[width = \linewidth]{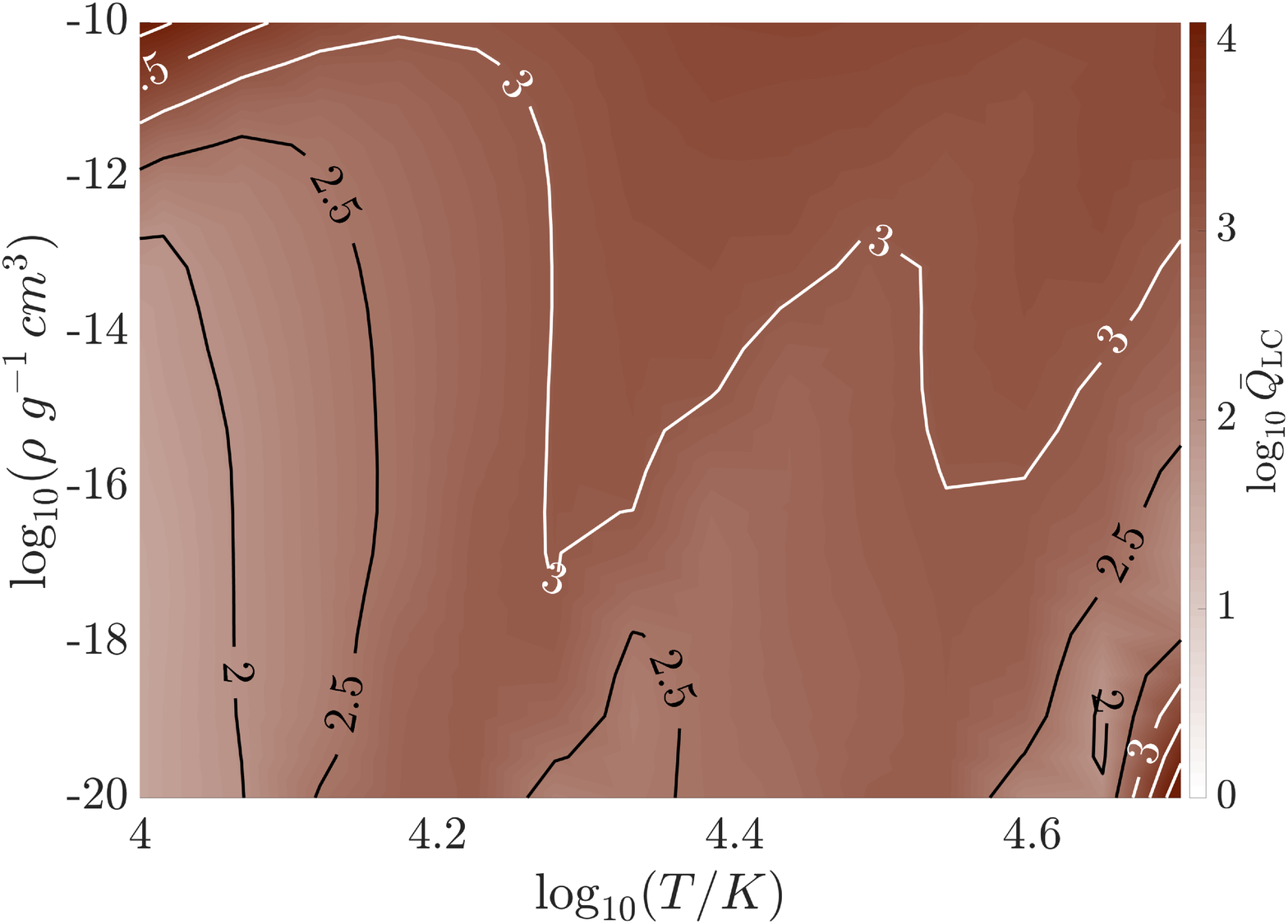}
	\caption{}
	\label{Fig_map_qb_lattimer}
	\end{subfigure}\\%
	\begin{subfigure}[t]{\linewidth}
		\includegraphics[width = \linewidth]{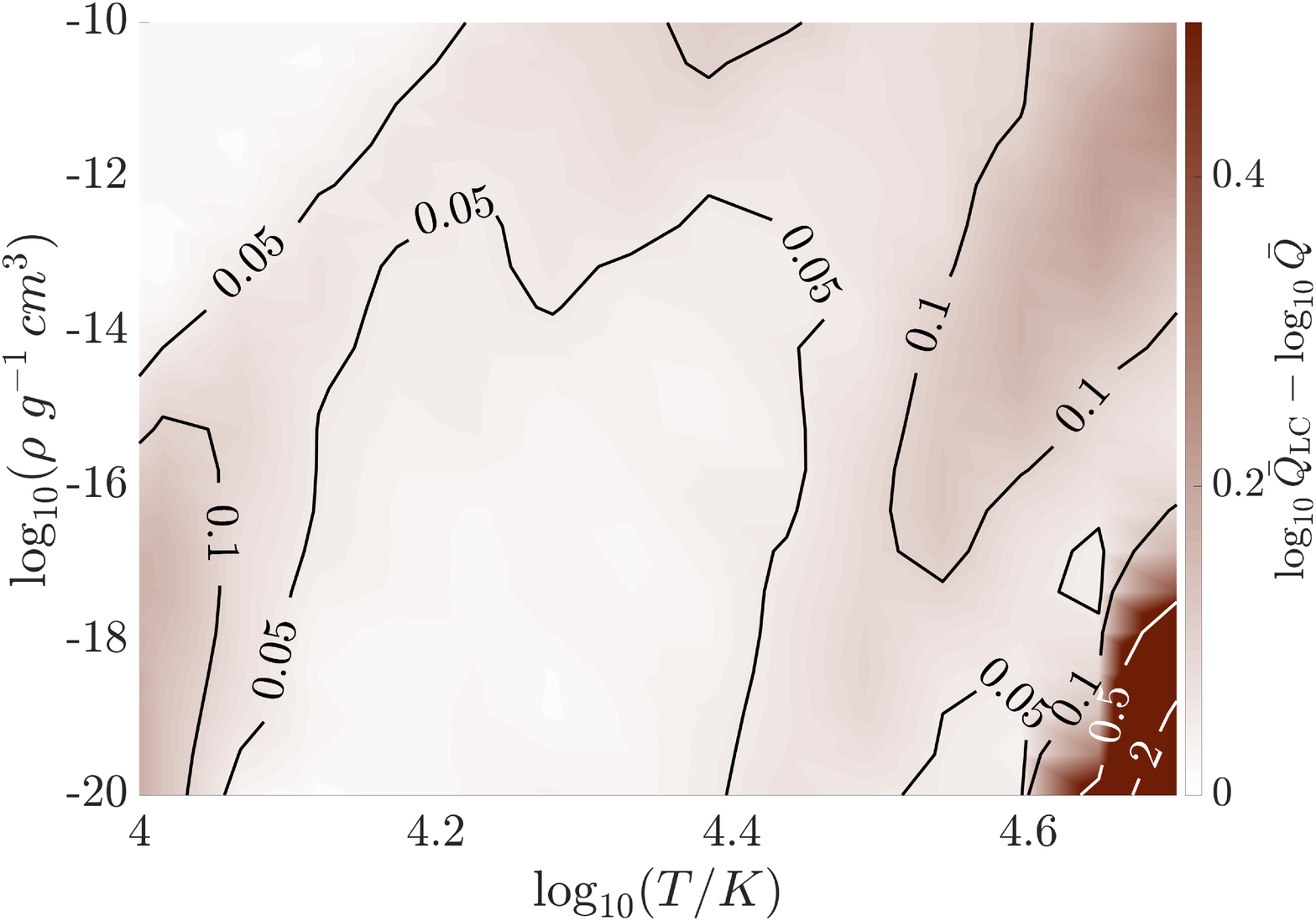}
	\caption{}
	\label{Fig_map_qb_lattimer_res}
	\end{subfigure}
	\caption{Colour maps of the $\log_{10}\bar{Q}_\lc$ computed by \citelc\ (a) and differences between the $\log_{10}\bar{Q}_\lc$ and $\log_{10}\bar{Q}$ presented on Fig.~\ref{Fig_map_qb} (b). Maps are given as a function of temperature on the abscissa and density on the ordinate. The $\log_{10}\bar{Q}_\lc$ is reconstructed from the data provided by \citelc.}
\end{figure}

Nonetheless, as our computations of the force multiplier are very similar, we may extract the corresponding $\bar{Q}$-values from the tabulations of \citelc\ in order to compare results stemming from different atomic line databases.
To achieve this, we first renormalise the results found by \citelc\ the same way as described in Sect.~\ref{Se_line_in_Sobolev} and then, in Fig.~\ref{Fig_map_qb_lattimer}, we plot the colour map of 
$\log_{10}\bar{Q}_\lc$ for a similar temperature and density grid as presented here\footnote{Data was obtained from the online repository provided by the authors \url{https://github.com/aslyv2/Rad-Winds}.}. 
For a quantitative comparison to our results (seen in Fig.~\ref{Fig_map_qb}), Fig.~\ref{Fig_map_qb_lattimer_res} then plots the ratio $\log_{10}(\bar{Q}_\lc/\bar{Q})$.

Inspecting these figures, we immediately see that the two independent results are broadly in agreement, with a small-factor difference throughout most of the parameter space. 
However, an exception is the low-density and high-temperature region, where the difference is significant.  
Since the only principal difference between our studies regards the line databases, this mismatch should stem from differences in these databases.

We note that the line database compiled by \citelc\ for elements includes ionisation stages up to ${\sc X}$ (i.e., nine times ionised), while the Munich line database (see Appendix~\ref{Se_atomic_data}) only extends to ionisation stages $\sc VIII$.
This difference in the available maximum ionisation stages becomes significant at low densities and high temperatures, as in this regime, it becomes increasingly easier to ionise elements. 
Therefore the significant difference between $\bar{Q}$ and $\bar{Q}_\lc$ in this region is most likely explained by the lack of spectral lines from high ionisation stages in our database. 
Nonetheless, we here focus on O-star applications, where the local conditions do not reach this corner of parameter space and thus should not pose an issue for our benchmark study. 
Indeed, examining the region of Fig.~\ref{Fig_map_qb_lattimer_res} corresponding to the density and temperatures of such O-star winds, the maximum difference between the two calculations is less than $20\%$ in $\bar Q$. 

\subsection{Limitations of the model}
\label{Se_dinc_limitations}
Having discussed the influence of the atomic data for our results, let us next focus on some of the limitations of the model. 
These limitations stem from the various approximations introduced in Sect.~\ref{Se_Flux_mean_opacity}, and here we focus on the three most important ones.

\subsubsection{NLTE effects and non-Planckian illumination}
\label{Se_disc_NonPlank}
 
In our work, we assume LTE and set the radiation and gas temperatures to be equal. 
Overall, as also discussed in Sect.~\ref{Se_Model-0}, this approximation seems to work reasonably well, resulting in O-star mass-loss rates that agree well with predictions derived from alternative models accounting for NLTE effects.
Nonetheless, NLTE effects are generally known to affect the detailed ionisation/excitation balance in hot stars, so a natural follow-up investigation to this study will be to examine effects from this also upon the detailed line-force tables presented here. 
As such, in a forthcoming paper, we will extend our method toward an approximate NLTE treatment (allowing the gas and radiation temperatures to be different) following, for example, the basic method outlined in \citet{Puls_00}.

In order to maintain generality when tabulating line-distribution parameters, we have approximated the frequency-distribution of the illuminating specific intensity with the Planck function. However, in a more realistic scenario, illumination may not be strictly Planckian, which might then affect the flux-weighted mean opacity.
In particular, if lines overlap within the velocity domain of an outflow, this can lead to multiple line-resonances, which may then affect the illumination of the individual resonance zones \citep[see, e.g.][]{Puls_93}. Moreover, if the gas is exposed to an
external strong ionising radiation field, as for example in active galactic nuclei (AGN) and high-mass X-ray binaries, this can further alter both the ionisation balance and the illuminating intensities \citep[see, e.g.,][]{Dannen_19}. 
Finally, by design, the straight line-force sum Eq.~\ref{Eq_Mt_sum} neglects all intrinsic line-overlap effects.
This may act towards reducing the line-force due to self-shadowing effects within intrinsically overlapping optically thick lines. 

\subsubsection{Non-Sobolev line force} 
\label{Se_disc_LDI}

In Sect.~\ref{Se_Power_law} we mentioned how a key advantage of the line-force parametrisation used here is that, it yields an explicit form for the underlying statistical distribution of lines. 
Using this statistical line-distribution function (i.e. Eq.~\ref{Eq_dN}), one can then also compute a corresponding non-Sobolev 
line force based on our tabulations, for example, by following the observer's frame escape-integral methods outlined in \citet{Owocki_96}. 
This would then allow us to investigate the effects from varying the line-force parameters upon the strong line-deshadowing instability \citep[LDI,][]{Owocki_84}, which is not captured within a Sobolev treatment. 
In terms of the general method developed here, a straightforward application would be to use our line-force tabulations directly in such LDI calculations. 
Then LDI models could be constructed without any (significant) further loss of computation time, and the effects of varying line-force parameters can be investigated in an analogous way as in this paper. 
Thus far, LDI simulations have always assumed prototypical line-force parameters (constant both in time and space), although it is clear that varying them will also impact the resulting LDI-generated structure \citep{Driessen_19}. 
First simulations of the LDI using self-consistent line-force parameters are already underway and will be presented in a forthcoming paper. 

\section{Conclusions}
\label{Se_conclusions}

This paper has presented a comprehensive tabulation of line-distribution parameters that can be used toward the fast calculation of flux-weighted line opacities and line forces in supersonic media (Sect.~\ref{Se_Flux_mean_opacity}). 
In Sect.~\ref{Se_first_appl}, we demonstrate that, when applied for (the rather well-calibrated) line-driven flows of O-stars our method (considering the limitations of the method outlined in Sect.~\ref{Se_dinc_limitations}) gives a very good agreement between our derived mass-loss rates and those obtained in the detailed (steady-state) studies by \citet{Krticka_17} and \citet{Bjorklund_21}. 
Based on this benchmark study, we here provide some important conclusions.

First, the self-consistent variation of the line-strength distribution parameters (i.e. $\alpha$, $\bar{Q}$, and $Q_0$) have critical feedback on the resulting line-driven flow structure, influencing the mass loss and velocity structure in the models.   
Thus we conclude that the conventional approach of keeping these parameters constant in time and space within multi-D, time-dependent line-driven flow simulations \citep[e.g.,][]{Kee_16, Dyda_18, Sundqvist_18, Schroder_21} may lead to inconsistent results.

Second, assuming LTE when computing the line-force seems to work reasonably well, at least in the near-star regions of O-stars (where their mass-loss rates are determined). Nonetheless, a key follow-up study to this paper will be an extension of the basic method presented here toward (approximate) NLTE conditions, allowing the gas and radiation temperatures to be different. 

%Third, our dynamical results suggests that requiring local dynamic consistency of the wind structure is vital when computing O-star mass-loss rates.
%In the other words, when neglecting the local consistency of the structure, one may arrive at a globally consistent solution that can not be locally sustained and thus is dynamically inconsistent \citep[see, e.g.,][where they find two globally consistent solutions with different velocity fields and mass-loss rates]{Gormaz_21}.

Finally, we conclude that the method presented here for computing flux-weighted line opacities is a beneficial, versatile, and very fast tool for implementing more realistic treatments of radiation line forces in multi-D, time-dependent RHD studies. 
For example, in a unified simulation encompassing both the optically thick, deep, and subsonic regions and an overlying line-driven wind, the method here can be trivially applied within the hybrid-opacity model proposed by \citet{Poniatowski_21} \citep[wherein $\kappa_{\rm F}$ is simply added to Rosseland mean opacities, see also discussion in][ch. 6]{Castor_04}. Indeed, using this hybrid opacity model, our new tables have already been implemented into the RHD code presented in \citet{Moens_21}, and first simulations of 3-D time-dependent WR wind outflows have been computed and will be presented in a parallel paper
\citep{Moens_22}.
%(Moens, Poniatowski, et al., in prep.). 

% (wherein $\kappa_{\rm line}$ is simply added to Rosseland mean opacities, see also discussion in Castor 2004, their Ch. 6)
%Finally, we provide a simple recipe for utilising this model in a general application. 
%As mentioned above, to better approximate opacity in the subsonic flows as well as in the optically thick media, we advise the reader to employ the hybrid opacity model \citep{Poniatowski_21}.
%Then the total opacity tensor will be the sum of the flux-weighted line opacities computed using Eq.~\ref{Eq_opacity_final} with the tabulated values of $\bar{Q}$, $Q_0$, and $\alpha$ and Rosseland mean opacities retrieved, for instance, using the OPAL opacity tabulation \citep{Iglesias_96}. \citep[see also discussion in][ch. 6]{Castor_04}

\begin{acknowledgements} 
The authors would like to thank all members of the KUL EQUATION group for fruitful discussion, comments, and suggestions.  
LP, JS, LD, AdK, and HS acknowledge support from the KU Leuven C1 grant MAESTRO C16/17/007.
JS further acknowledges additional support by the Belgian Research Foundation Flanders (FWO) Odysseus program under grant number G0H9218N. NDK acknowledges support from the National Solar Observatory (NSO), which is operated by the Association of Universities for Research in Astronomy, Inc. (AURA), under cooperative agreement with the National Science Foundation. We finally also thank the referee for their comments on the manuscript.
\end{acknowledgements}

%%%%%%%%%%%%%%%%%%%%%%%%%%%%%%%%%%%%%%%%%%%%%%%%%%%%%%%
%%%%%%%%%%%%%%%%%%%%%%%%%%%%%%%%%%%%%%%%%%%%%%%%%%%%%%%
%%%%%%%%%%%%%%%%%%%%%%%%%%%%%%%%%%%%%%%%%%%%%%%%%%%%%%%
\appendix
%%%%%%%%%%%%%%%%%%%%%%%%%%%%%%%%%%%%%%%%%%%%%%%%%%%%%%%
%%%%%%%%%%%%%%%%%%%%%%%%%%%%%%%%%%%%%%%%%%%%%%%%%%%%%%%
%%%%%%%%%%%%%%%%%%%%%%%%%%%%%%%%%%%%%%%%%%%%%%%%%%%%%%%

\section{Ionisation and excitation balance}\label{Se_balance}

In this section, we describe how we find the ionisation and excitation balance required for the computation of the line force in Sect.~\ref{Se_discrete_sum}.
Under the LTE assumption, the ionisation balance is computed using the Saha-Boltzmann equations.
We obtain the ionisation balance by first computing the number density $n_{z}$ of a specific element with nuclear charge $z$, 
and atomic weight $A_z$ (for hydrogen $A_z = m_H$). 
Taking the number abundance relative to hydrogen $N_z$ \citep[relative number abundances in this paper are obtained from the tabulated solar values in][]{Asplund_09} and the mass fraction of hydrogen:
\begin{equation}
X = \frac{m_H}{\sum_{z} A_z\; N_z}\,,
\end{equation}
the number density of element $z$ for a plasma with mass density $\rho$ is:
\begin{equation}
n_z = \rho\frac{N_z X}{m_H}\,.
\end{equation} 

Partition functions of every ionisation stage $i$ of each element $z$ are computed as:
\begin{equation}
U_{z\; i} = \sum\limits_{l = 1}^{l_{\max}} g_l \exp \left( \frac{\epsilon_{z\; i} - \varepsilon_{z\; i\; l} }{\kb T}\right)\, ,
\end{equation}
where $g_l$ is the statistical weight of level $l$, $\epsilon_{z\; i}$ is the ionisation energy from the ground level, 
and $\varepsilon_{z\; i\; l}$ is the excitation energy of the corresponding level.
% (for ground level $\varepsilon_{z,\,i,\,l} = \epsilon_{z,\,i}$).
In general, sums are taken over all levels; however, as higher excitation levels generally have a small contribution, we only sum to some highest excitation level $l_{\max}$ (see Appendix~\ref{Se_atomic_data} for details on the atomic data used in this paper, including information on maximum ionisation states and excitation levels for each ion).
Since we have access to the extensive Munich atomic database, we opt to compute the partition functions using the above sum throughout our work, (rather than using fit functions such as those provided by \citealt{Cardona_10}). 
This is the same way as is being done within the well-calibrated, standard 1-D model atmosphere code \texttt{FASTWIND} \citep{Puls_05}, which also uses the same atomic database as here.

Now the Saha equation is used to compute the number densities of ionisation stage $i+1$ relative to $i$ for each element, 
\begin{equation}\label{Eq_saha}
\frac{n_{z\; i+1}}{n_{z\; i}} = \frac{2 U_{z\; i+1}}{n_e U_{z\; i}} 
\left(\frac{2\pi m_e \kb T}{h^2}\right)^{\sfrac{3}{2}} 
\exp\left(\frac{\epsilon_{z\;i} - \epsilon_{z\; i+1}}{\kb T}\right)\,.
\end{equation}
Here $n_e$ is the number density of free electrons, and  $n_{z\, i}$ and $n_{z\, i+1}$ are the number densities of the $i$ and $i+1$ ionisation stages of element $z$. 
It is further convenient to express number densities for each ionisation stage with respect to the neutral stage of the element:
\begin{equation}\label{Eq_ion_prod}
\frac{n_{z\; i}}{n_{z\, 1}} = \prod\limits_{j=1}^{i-1}\frac{n_{z\; j+1}}{n_{z\; j}}\, ,
\end{equation}
where fractions on the right-hand side of expression are given by Eq.~\ref{Eq_saha}.
Knowing $n_{z\; 1}$ the expression above gives us the number density of each ionisation stage.  
To find $n_{z\; 1}$, we simply apply conservation of total number density:
\begin{equation}\label{Eq_ion_sum}
n_{z\; 1} = n_z\left(\sum\limits_{i=1}^{i_{\max}} \frac{n_{z\; i}}{n_{z\; 1}}  \right)^{-1}\, ,
\end{equation}
where the fractions in the sum are given by Eq.~\ref{Eq_ion_prod}.

Finally, it is important to note that these calculations depend on the number density of free electrons $n_e$, which in turn depends on the result of the same calculations through:
\begin{equation}\label{Eq_ne}
n_e = \sum\limits_{z=1}^{z_{\max}}\sum\limits_{i=1}^{i_{\max}} (i-1)n_{z\; i}\, .
\end{equation}
As such, iteration is generally required to determine the correct ionisation balance. In the first step of the iteration, $n_e$ is set by assuming fully ionised hydrogen. 
%the rest of the elements contribute half of their electrons. 
This then allows to solve for the ionisation balance and compute new $n_e^{\rm new} $ using Eq.~\ref{Eq_ne}. 
We then update the electron number density in the next iteration by taking the square root of the product of its values from the current and previous iterations \citepalias[see also][]{Lattimer_21}.

Now applying the Boltzmann excitation law, the occupation number density of element $z$ in excitation level $l$ and ionisation state $i$ with respect to the ground-state occupation number density $n_{z\, i\,1}$ is:
\begin{equation}
\label{Eq_Boltzmann}
\frac{n_{z\; i\; l}}{n_{z\; i\; 1}} = \frac{g_l}{g_1}\exp\left(\frac{\epsilon_{z\; i} - \varepsilon_{z\; i\; l}}{\kb T}\right)\,.
\end{equation}
Again applying conservation of number density, we get:
\begin{equation}
\label{Eq_Boltzmann_sum}
n_{z\; i\; l}= n_{z\; i}\;\frac{n_{z\; i\; l}}{n_{z\; i\; 1}} \left(\sum\limits_{l = 1}^{l_{\max}} \frac{n_{z\; i\; l}}{n_{z\; i\; 1}} \right)^{-1}\,,
\end{equation}
where the fraction terms come from Eq.~\ref{Eq_Boltzmann}.
In this way, we compute atomic occupation number densities for each excitation level of every ion contained in our database.

\section{Atomic data}
\label{Se_atomic_data}

For the study presented in this paper, we used the Munich atomic database originally compiled by \citet{Pauldrach_98, Pauldrach_01}, see also \citet{Puls_05}. 
This database contains elements from hydrogen to zinc, with the exception of Li, Be, B, and Sc, which are rare and provide minimal effect. 
For the included elements, the database provides ionisation stages up to VIII. The total number of lines included is $4.2\cdot10^{6}$, for which $gf$-values of their lower levels are provided. 
We provide a detailed list of available ionisation stages for the element and the number of available lines for each ionisation stage in Table~\ref{Tbl_atomic_data}.

\begin{table*}
\centering
\caption{The list of available ionisation stages for elements form H to Zn, and the number of available lines for each ionisation stage included in the database used in our study.}
\label{Tbl_atomic_data}
\begin{tabular}{l  l l l  l l l  l l  }
\hline\hline
\multirow{2}{*}{Element} & \multicolumn{8}{c}{Number of lines per ion} \\ 
                                        & I & II & III & IV & V & VI & VII & VIII \\ 
\hline
 H    & 39     & -       & -       &  -     & -      &-         & -        & -\\
 He   & 113    & 82      & -       & -      & -      & -        & -        &- \\
 Li   & -      &-        &  -      &  -     &  -     & -        &  -       &- \\
 Be   & -      &  -      &    -    &     -  &     -  &       -  &        - &- \\
 B    & -      &    -    &     -   &   -    &    -   &      -   &     -    &- \\
 C    & 1330   & 11005   & 4406    & 229    & 57     & -        &   -      &- \\
 N    & 2142   & 747     & 16458   & 4401   & 229    & 57       & -        &- \\
 O    & 18315  & 39207   & 24506   & 17933  & 4336   & 231      & -        &- \\
 F    & 853    & 607     & 409     & 198    & 144    &16        & -        &- \\
 
 Ne   & 584    & 2564    & 857     & 4470   & 2664   & 1912     & -        & -\\
 Na   & 84     & 255     & 1040    & 34     & 53     & 45       &  -       &- \\
 Mg   & 432    & 3371    & 2457    & 3669   & 3439   & 305      &  -       &- \\
 Al   & 99     & 924     & 273     & 2523   & 18317  & 153      &  -       &- \\
 Si   & 1363   & 4560    & 4044    & 245    & 3096   & 3889     &  -       & -\\
 P    & 560    & 405     & 139     & 87     & 245    & 1096     & -        & -\\
 S    & 281    & 27767   & 190     & 70     & 903    & 142      & 1031     &-\\
 Cl   & 1085   & 705     & 432     & 132    & 73     & 74       &  -       & -\\
 Ar   & 870    & 1967    & 398     & 150    & 3007   & 1335     & 2198     & 111\\
 K    & 156    & 63      & 188     & 65     & 138    & 16       & -        & -\\
 Ca   & 643    & 3067    & 636     & 20030  & 28685  & 19       & -        & -\\
 Sc   & -      & -       &-        &     -  & -      &     -    &     -    &- \\

 Ti   & 2801   & 56681   & 8599    & 48     & 4      & -        & -        &- \\
 V    & 3328   & 2385    & 976     & 560    & 18     & -        & -        & -\\
 Cr   & 4105   & 221932  & 134687  & 60264  & 88     & 5955     &  -       & -\\
 Mn   & 2085   & 227761  & 175593  & 131821 & 61790  & 87       & -        &- \\
 Fe   & 3770   & 227548  & 199484  & 172902 & 124157 & 60458    & 10123    & 4777\\
 Co   & 2329   & 189858  & 200637  & 146252 & 182780 & 124053   & 50270    & -\\
 Ni   & 1734   & 129790  & 131508  & 183267 & 179921 & 186055   & 123386   & 43778\\
 Cu   & 166    & 1133    &  490    & 17466  & 30457  & 10849    & -        & -\\
 Zn   & 16     & 2       & 1194    & -      &     -  &        - &    -     &- \\
\hline
\end{tabular}
\end{table*}

\bibliographystyle{aa} % style aa.bst
\bibliography{biblio_lnk} % your references Yourfile.bib
\end{document}